\documentclass[11pt,twocolumn]{emulateapj}

\tightenlines
\usepackage{epstopdf}
\usepackage{ulem} 
\usepackage{color} 
\usepackage{lineno}
\usepackage{amsmath}

\pdfoutput=1

\def\lsim{\;\raise0.3ex\hbox{$<$\kern-0.75em\raise-1.1ex\hbox{$\sim$}}\;}
\def\gsim{\;\raise0.3ex\hbox{$>$\kern-0.75em\raise-1.1ex\hbox{$\sim$}}\;}

\definecolor{purple}{RGB}{200,100,255} 

\newcommand{\dNdp}{dN/dp}
\newcommand{\Pinj}{P_\mathrm{inj}}

\newcommand{\Ng}{N_g}
\newcommand{\delthtmax}{\delta\theta_\mathrm{max}}

\newcommand{\xx}[1]{\!\times\!10^{#1}}

\newcommand{\pcc}{cm$^{-3}$}

\newcommand{\muG}{$\mu$G}
\newcommand{\RH}{Rankine--Hugoniot}
\newcommand{\rRH}{R_\mathrm{RH}}

\newcommand{\rg}{r_g}
\newcommand{\rgZ}{r_{g0}}
\newcommand{\epse}{\epsilon_{e}}
\newcommand{\epsB}{\epsilon_{B}}

\newcommand{\lambdamfp}{\lambda_\mathrm{mfp}}

\renewcommand{\baselinestretch}{1}
\def\aap{Astronomy and Astrophysics}

\def\apjl{The Astrophysical Journal Letters}

\newcount\listnorom
\newcommand\listromanDE{\global\advance \listnorom by 1
{\lowercase\expandafter{(\romannumeral\listnorom)}\ }}

\newcount\listnumber
\newcommand\listDE{\global\advance \listnumber by 1
{\lowercase\expandafter{(\number\listnumber)}\ }}

\newcount\Inum
\newcount\IInum
\newcount\IIInum
\newcount\IVnum
\def\I{\global\multiply\IInum by 0 \global\multiply\IIInum by 0
            \global\multiply\IVnum by 0 \global\advance \Inum by 1
            {\the\Inum. }}
\def\II{\global\multiply\IIInum by 0\global\multiply\IVnum by 0
       \global\advance \IInum by 1 {\the\Inum.\the\IInum. }}
\def\III{\global\multiply\IVnum by 0\global\advance \IIInum by 1
            {\the\Inum.\the\IInum.\the\IIInum. }}
\def\IV{\global\advance \IVnum by 1
            {\the\IVnum. }}
\shorttitle{NLDSA model for GRB afterglows}
\shortauthors{Warren et al.}

\begin{document}

\title{Nonlinear particle acceleration and thermal particles in GRB afterglows} 

\vskip24pt

\author{Donald C. Warren,\altaffilmark{1}
Donald C. Ellison,\altaffilmark{2}
Maxim V. Barkov,\altaffilmark{1,3,4}
Shigehiro Nagataki\altaffilmark{1,5}}

\altaffiltext{1}{Astrophysical Big Bang Laboratory, RIKEN, Saitama 351-0198, Japan; donald.warren@riken.jp}

\altaffiltext{2}{Physics Department, North Carolina State
University, Box 8202, Raleigh, NC 27695, U.S.A.}

\altaffiltext{3}{Institute for Physics and Astronomy, University of Potsdam, 11476 Potsdam, Germany}

\altaffiltext{4}{DESY, Platanenallee 6, 15378 Zeuthen, Germany}

\altaffiltext{5}{Interdisicplinary Theoretical Science Research Group (iTHES), RIKEN, Saitama 351-0198, Japan}

\begin{abstract}

The standard model for GRB afterglow emission treats the accelerated electron population as a simple power law, $N(E) \propto E^{-p}$ for $p \gtrsim 2$.  However, in standard Fermi shock acceleration a substantial fraction of the swept-up particles do not enter the acceleration process at all.  Additionally, if acceleration is efficient then the nonlinear backreaction of accelerated particles on the shock structure modifies the shape of the non-thermal tail of the particle spectra.  Both of these modifications to the standard synchrotron afterglow impact the luminosity, spectra, and temporal variation of the afterglow.  To examine the effects of including thermal particles and nonlinear particle acceleration on afterglow emission, we follow a hydrodynamical model for an afterglow jet and simulate acceleration at numerous points during the evolution.  When thermal particles are included, we find that the electron population is at no time well-fitted by a single power law, though the highest-energy electrons are; if the acceleration is efficient, then the power law region is even smaller.  Our model predicts hard-soft-hard spectral evolution at X-ray energies, as well as an uncoupled X-ray and optical light curve.  Additionally, we show that including emission from thermal particles has drastic effects (factors of 100 and 30, respectively) on the observed flux at optical and GeV energies.  This enhancement of GeV emission makes afterglow detections by future $\gamma$-ray observatories, such as CTA, very likely.

Keywords: acceleration of particles --- ISM: cosmic rays --- gamma-ray bursts --- magnetohydrodynamics (MHD) --- shock waves  --- turbulence
\end{abstract}

\section{Introduction}
\label{sec:intro}

Gamma-ray bursts (GRBs) are brief flashes of gamma-ray photons occurring a few times a day across the entire sky.  After the high-energy prompt emission fades, broadband (usually radio to X-ray, but occasionally GeV gamma-ray) afterglow emission is sometimes observed.  These afterglows are less numerous, though increasingly frequent as technology improves: the \textit{Swift} mission \citep{Gehrels_etal_2004} alone reports over 300 GRBs with afterglows detected both in optical/UV and in X-ray\footnote{http://swift.gsfc.nasa.gov/archive/grb\_table/stats/}, and of the $\sim 1800$ GRBs reported between 1997 and 2015\footnote{http://www.astro.caltech.edu/grbox/grbox.php} more than 1000 were detected in at least one of radio, optical, or X-ray---with 84 detected in all three\footnote{http://www.mpe.mpg.de/~jcg/grbgen.html}. Afterglow observations offer a wealth of information about the progenitors of GRBs, their environments, the microphysics of particle acceleration, and more. A full accounting of the physics of afterglows is beyond the scope of this work, but see \citet{KumarZhang2015} for an extended review of the current state of the field.  Suffice it to say that correctly interpreting observations of afterglows is critical to understanding the physics going on in them, and that a great deal of work has been done in service of this goal.

The traditional picture of GRB afterglows is that they are synchrotron emission from a population of energetic electrons accelerated by a relativistic external shock. This model for afterglow emission predates the first afterglow detection \citep{PaczynskiRhoads1993,MeszarosRees1997}.  Early analytical models treating the time-dependence of the afterglow invoked synchrotron radiation from a single power law of electrons \citep{SPN1998,ChevalierLi2000}.  Though very efficient in terms of computation time, these models are heavily dependent on assumptions about, e.g., the magnetohydrodynamic state of the fluid in the vicinity of the blast wave, the presence of a reverse shock, whether the environment of the GRB was wind-like or of constant density, and many more.  The numerous permutations, and their consequences for both single-time spectra and long-term light curves, are collected in \citet{Gao_etal_2013}.  Numerical studies have also been undertaken.  Numerical hydrodynamics allows more accurate handling of emission (1) during the trans-relativistic phase of shock deceleration \citep{vanEerten_etal_2010}, (2) from a reverse shock \citep{Uhm_etal_2012}, (3) due to the multi-dimensional nature of the blast-wave \citep{vEvdHMF2012,vanEertenMacFadyen2012}, and in many other scenarios.

Despite the multitude of theoretical and numerical analyses and the breadth of situations considered, virtually all work on afterglow emission shares one common assumption: that the electrons responsible for the broadband emission form a power law with a single spectral index $p$ (this letter will also be used throughout this paper to refer to momentum, but context should make it clear which definition applies at any particular instance).  The rare exceptions considered a Maxwellian thermal distribution instead of \citep{PVP2014} or in addition to \citep{GianniosSpitkovsky2009} a power law tail.\footnote{The work of \citet{EichlerWaxman2005} also treated the possibility of separate thermal and accelerated electron populations, but used unreasonably low values for the thermal electron energy because crucial numerical results had not yet been published.}

The value for the electron spectral index $p$, starting from basic shock acceleration principles, is 2.23 \citep{Kirk_etal_2000,KeshetWaxman2005}, in agreement with the results of Monte Carlo simulations \citep{ERJ1990,BednarzOstrowski1998,Achterberg_etal_2001,EWB2013}.  Strictly speaking, this result applies only for ultra-relativistic shocks in the case where the upstream magnetic field and shock normal are parallel, and then only in the limit where accelerated particles carry negligible energy (the so-called ``test particle'' regime).  However, particle-in-cell (PIC) simulations show that self-generated magnetic field turbulence can overwhelm weak pre-existing fields (those for which the magnetization parameter of the inflowing plasma is $\sigma \lesssim 10^{-5}$, with $\sigma \equiv B_{0}^{2}/(4\pi\gamma_{0}n_{0}m_{p}c^{2})$ for a shock moving with Lorentz factor $\gamma_{0}$ into plasma with mean magnetic field $B_{0}$ and proton number density $n_{0}$) at relativistic shocks \citep{SSA2013}, and the results are very similar to those at parallel or entirely unmagnetized shocks \citep[e.g.,][]{Spitkovsky2008}.

Observations of GRB afterglows suggest softer spectra than predicted by test-particle theory.  \citet{Curran_etal_2010} attempted to fit X-ray observations both to a single universal power law and to a distribution of $p$ values.  Their most likely single value for $p$ was 2.25, in close agreement with theory, but statistical tests rejected a universal $p$ value.  When a Gaussian distribution of spectral indices was used instead, the best fit was $p = 2.36 \pm 0.59$.  The large uncertainty is almost entirely due to a long tail of softer---rather than harder---particle spectra.  A search over a larger parameter space (varying, e.g., explosion energy, electron spectral index, jet opening angle, and energy density in electrons and magnetic field) found that GRB afterglows spanned the entire considered range of spectral indices, as soft as $p = 5$ \citep{Ryan_etal_2015}.  For lack of a universal value for $p$ stemming from observations, the traditional value used in studies of GRB afterglows is $p = 2.4-2.5$.

It is well known that at non-relativistic speeds, accelerated particles may not form a simple power law.  Efficient acceleration of particles\footnote{We must be clear, here: if even 5\% of the kinetic energy of inflowing particles (viewed in the frame where the shock is at rest) is transferred to a high-energy distribution of accelerated particles, then the shock is beyond the test-particle regime.  This is lower than the $\sim10\%$ observed in long-term electron-ion PIC simulations \citep{SSA2013}.} moves such shocks beyond the test-particle regime, as the accelerated particles influence the structure of the shock doing the accelerating.  This nonlinear feedback loop results in a shock precursor, which acts to slightly slow the inflowing plasma before it crosses the shock.  More energetic particles thus experience a larger velocity difference between the upstream and downstream limits of their scattering, which in turn causes a harder spectrum at higher particle energies \citep[see, e.g., discussion in][]{BerezhkoEllison1999,BGV2005}.

While most work on nonlinear Fermi acceleration has been done for ions accelerated in nonrelativistic shocks, it is important to emphasize that nonlinear effects must occur in relativistic shocks as well, if acceleration is efficient.  Furthermore, if no mechanism acts to transfer energy from ions to electrons, Fermi acceleration places a large majority of the shock energy into ions.  \citet{EWB2013} have shown that the nonlinear effects required to ensure flux conservation in relativistic shocks may lead to orders of magnitude depression of electron acceleration efficiency \citep{WEBL2015}.  Handling the interaction between shocks and the particles they accelerate is of paramount importance to properly understanding GRB afterglows.

In addition to the electron spectral index $p$, GRB models use two parameters to handle the microphysics present at the shock: the fraction of energy density in electrons ($\epse$; see discussion surrounding Equation~\ref{eq:epse}) and magnetic field ($\epsB$).  Applying the standard synchrotron model for afterglows shows that the value of $\epse$ in most bursts is $0.1-0.3$ \citep{SBDK2014}.  The value of $\epsB$ is much less constrained and may vary over several orders of magnitude, from $10^{-7}$ to $10^{-3}$ \citep{SBDK2014,Wang_etal_2015}.  If a particularly hard electron spectral index $p \lesssim 2.4$ is chosen, $\epsB$ may even approach $10^{-2}$.  These ranges are significantly larger than the standard interstellar medium (ISM) value of $\epsB \sim 10^{-9}$, and require amplification of the magnetic field by means of plasma instabilities.

All of the preceding discussion of $\epse$ and $\epsB$ rests on the assumption that the two parameters are constant throughout the life of the afterglow.  This is the default state for afterglow modeling, though alternatives have been explored in a handful of cases.  \citet{RossiRees2003} explored a basic two-zone model for the external shock and afterglow, allowing for different conditions at and downstream from the shock.  Letting $\epse$ and $\epsB$ vary with time has been proposed to explain both the plateau phase of many X-ray light curves \citep{Ioka_etal_2006} and the chromatic spectral breaks seen in many afterglows \citep{Panaitescu_etal_2006MNRAS369}.  The two parameters have been fitted to decaying power laws in order to explore rebrightenings associated with a wind-ISM interface \citep{Kong_etal_2010}.  The possibility that $\epse$ and $\epsB$ might \textit{increase} with time was considered in \citet{GKP2006}, as a means to reduce the $\gamma$-ray efficiency of the prompt emission below the worryingly large values of $0.5-0.9$.  In each of these cases, however, the values for $\epse$ and $\epsB$ were either chosen to match observations or given a specific functional form.

No work of which we are aware has calculated $\epse$ or $\epsB$ from first principles over a large portion of an afterglow.\footnote{PIC simulations such as \citet{SSA2013} and \citet{Ardaneh_etal_2015}, of course, allow for direct calculation of both $\epse$ and $\epsB$.  However, these simulations are limited by computational power to studying time periods vastly shorter than the timescales of afterglows.}  In \citet{Lemoine2013}, though, a functional form for decaying microturbulence was assumed based on PIC results.  The synchrotron spectrum of particles downstream from the shock was calculated using that functional form in a variety of conditions (e.g., gradual/rapid decay of turbulence, weak/strong inverse Compton losses, etc.).  The longitudinal extent of the enhanced turbulence was some $10^{2}-10^{4}$ ion skin depths.  While this is a large distance for PIC simulations, it is ignorably small compared to the volumes being simulated in this work.  The length scale of our Monte Carlo code is $\rgZ = \gamma_{0}u_{0} m_{p} c / (eB_{0})$, which is $\sigma^{-1/2}$ times larger than the plasma skin depth; the emission shells are individually thousands of $\rgZ$ in size.  Here, $u_{0}$ is the shock speed, $m_{p}$ is the proton mass, $c$, and $e$ is the electronic charge.

The aim of this paper is to build a physically-motivated model of a GRB afterglow, from early in the afterglow until very nearly the transrelativistic stage of deceleration.  We pay particular attention to the nonlinear effects, from the simultaneous acceleration of electrons and protons, linking the shock structure and the cosmic rays being accelerated.  Where needed, we incorporate results from PIC simulations to guide our parametrization of key microphysics.  We calculate photon spectra for various times, which we compare against observations in the optical, X-ray, and gamma-ray bands.

The paper is organized as follows.  In \S\ref{sec:MCcode} we discuss the Monte Carlo code used to generate the particle and photon spectra to be presented later.  The Monte Carlo code is applied post-process to a hydroynamical base to create a sequence of afterglow snapshots, in a manner we describe in \S\ref{sec:evomodel}.  We present our example afterglows and compare against observations in \S\ref{sec:afterglows}, and conclude in \S\ref{sec:conclusions}.

\section{The Monte Carlo code}
\label{sec:MCcode}

The work presented here relies on a Monte Carlo code treating first-order Fermi acceleration in the vicinity of an infinite, steady-state, parallel, collisionless shock.  The code follows the scattering and propagation of particles in a turbulent magnetic field around a velocity gradient.  Momentum and energy fluxes are tracked during propagation, which may then be used to modify the shock structure in an attempt to satisfy the \RH\ conditions at all locations.  The process has been described in detail in, e.g., \citet{EWB2013} and \citet{WEBL2015}, and the reader is referred to those papers for a more complete description of the code.  In this section we briefly discuss key features, and departures from previous work.

All particles---regardless of energy or species---are assumed to scatter elastically off turbulence in the local fluid frame with a mean free path given by
\begin{equation}
  \lambdamfp = \rg ,
  \label{eq:lambdamfp}
\end{equation}
where $\rg = pc/ (ZeB)$ is the gyroradius of a particle with charge number $Z$ in the local magnetic field $B$.  That is, we assume that particle scattering occurs in the Bohm limit.  This assumption fails in the immediate vicinity of a relativistic shock, as the magnetic field is turbulent down to length scales of the ion skin depth, which may be many orders of magnitude smaller than the particle gyroradius \citep[e.g.,][]{SSA2013,PPL2013}.  In these conditions the mean free path approaches $\lambdamfp \propto p^{2}$ rather than $\propto p^{1}$ \citep[e.g.,][]{PPL2011}.  If, on the other hand, the power spectrum of the magnetic turbulence extends to length scales comparable to $\rg$, then scattering proceeds as $p^{1}$ at all momenta.  We do not attempt to address these complexities at this time; previous work on this topic has shown that this simple model for particle scattering can adequately reproduce some important PIC results for injection and early acceleration \citep{EWB2013}.  \citet{EWB2016} discussed the effect of a modified diffusion coefficient on both CR spectra and nonlinear smoothing, and we refer the reader to that paper for more information.

For now we simply note that the region of intense turbulence is contained to a small region around the shock, and so we assume that the effects of this region on the overall model are minimal.  In the upstream direction the small-scale Weibel instability is contained to a precursor extending a short distance ahead of the shock \citep[see Equation~6 in][]{SSA2013}.  Downstream from the shock the turbulence decays rapidly, inverse-cascading to larger wavelengths and decreasing in strength as $t^{-1/2}$ \citep[][and references therein]{LLW2013}.  Both scales are dramatically smaller than the distances considered in this paper.

We assume pitch-angle scattering for the particles as they move in the turbulent magnetic field.  Under the pitch-angle scattering model, a particle with velocity $v_\mathrm{pt}$ travels for a time $\delta t \ll t_{c}$ between scatterings.  The quantity $t_{c} = \lambdamfp / v_\mathrm{pt}$ is the ``collision time,'' the fluid-frame time to accumulate a cumulative deflection on the order of $90^{\circ}$.  Our implementation of pitch-angle scattering requires a parameter, $\Ng$, which controls the fineness of scattering.  Each gyroperiod $\tau_{g} = 2\pi \rg / v_\mathrm{pt}$ is divided into $\Ng$ time steps, so $\delta t = \tau_{g} / \Ng$.  The maximum deflection a particle may experience at a single scattering event is also controlled by $\Ng$ \citep{ERJ1990},\footnote{This equation is specific to the choice of $\lambdamfp$ made in Equation~\ref{eq:lambdamfp}.  More generally, if a particle's mean free path is a function $f(\rg)$ of its gyroradius, Equation~\ref{eq:delthtmax} may be rewritten $\delthtmax = \sqrt{12 \pi \rg / (\Ng f(\rg))}$.}
\begin{equation}
  \delthtmax = \sqrt{12 \pi / \Ng} .
  \label{eq:delthtmax}
\end{equation}
Large values of $\Ng$ imply that only modest deflections occur at scattering events, while small values of $\Ng$ suggest large-angle scattering \citep{ERJ1990,SBS2007,SummerlinBaring2012}.

The code does not attempt to calculate the wave--particle interactions that govern particle motion in physical shocks.  Equations~\ref{eq:lambdamfp} and \ref{eq:delthtmax} control the transport of particles within the shock structure.  As such, we assume that sufficient turbulence exists on the correct length scales to drive that transport.  Further, note that both equations are position-independent: though position-dependent diffusion coefficients have been incorporated in Monte Carlo calculations in the nonrelativistic regime \citep{VBE2009,Bykov_etal_2014}, we defer that particular extension of our code to future work.

The assumption of position-independent scattering, even in the upstream region, is somewhat at odds with theoretical predictions \citep{LemoinePelletier2010} and PIC simulations of relativistic shocks propagating into moderately strong upstream magnetic fields \citep{SSA2013}.  (In the simulations of the latter work, computational costs precluded testing the result at the very low upstream magnetic field energy fractions expected in GRB afterglow shocks.)  In both of those works, the longitudinal extent of the turbulent upstream precursor is a few times $\rgZ$ ($\equiv \gamma_{0} m_{p} u_{0} c / (eB_{0})$, the gyroradius of a proton with speed $u_{0}$ in a field of strength $B_{0}$).  When the upstream magnetic field is ordered rather than turbulent, the upstream dwell time of particles increases by a factor of at least a few \citep{Achterberg_etal_2001}.  For CR acceleration happening over a given (and limited) time period, then, the turbulent-field case allows for fewer shock crossing cycles, as particles can scatter further upstream from the shock every time they cross back into the upstream region.  Protons, which do not experience any sort of loss process in our simulation, are limited in maximum energy entirely by acceleration time.  Our assumption of a turbulent upstream region therefore suppresses the maximum energy of the proton spectrum, and there is good reason to suspect that the CR tail would extend to higher energies with a more exact model of the magnetic field around the shock.  Electrons, on the other hand, are limited by synchrotron losses incurred in the downstream region; their spectrum is unaffected by turbulence upstream of the shock.

Particle injection is handled via a thermal leakage model.  As unshocked particles first cross the shock and interact with the downstream plasma, they will scatter in the downstream plasma frame and be energized.  Since the downstream flow is subsonic, this ``thermalization'' results in a significant fraction of particles having a plasma frame speed greater than the bulk fluid speed, which is only mildly relativistic regardless of $\gamma_{0}$.  The Monte Carlo code follows these particles as they scatter back across the subshock into the upstream region.  A particle that re-enters the upstream region will be further accelerated as it scatters back downstream, which is our definition of injection.  Implicit in this definition is the assumption that the subshock is transparent; in other words, mildly superthermal particles can cross without experiencing forces from large-amplitude magnetic turbulence, cross-shock potentials, or other phenomena that may influence the injection rate.

We do not pretend that this injection captures all of the complicated plasma physics seen by PIC simulations of the shock layer.  Instead, we parametrize the injection probability in a straightforward fashion and connect the injection self-consistently to Fermi acceleration and the nonlinear shock structure.  It is also trivial to include the simultaneous injection and acceleration of electrons in addition to protons.  Since Monte Carlo transport allows for arbitrary particle anisotropy, it has unique advantages over semi-analytic methods that avoid direcly modeling thermal particles.  Furthermore, the simple parametrization of complex plasma processes allows modeling Fermi acceleration, with internally consistent injection, to energies not yet accessible with PIC simulations.

For the basic scattering assumptions, i.e. Equations~\ref{eq:lambdamfp} and \ref{eq:delthtmax}, this injection model generally predicts efficient injection and acceleration incompatible with the test particle limit \citep[e.g.,][]{WEBL2015}.  If a sharp, unmodified shock is assumed, then energy and momentum fluxes are not conserved within the shock structure: the flux downstream from the shock will always exceed the incoming flux from far upstream.

In order to discuss test-particle shocks, we have introduced a new parameter, $\Pinj \in [0,1]$.  When a thermal particle downstream from the shock would recross the shock, $\Pinj$ is the chance that it will be injected into acceleration process.  If the particle fails that check, it is reflected back downstream to continue scattering.  Note that $\Pinj = 0$ corresponds to disabling all diffusive shock acceleration \citep[e.g. Figure~1 in][]{EWB2013}, while $\Pinj = 1$ is the thermal leakage model already mentioned.

For any value of $\Pinj$, the code can determine the self-consistent, modified shock structure, including the back-reaction of accelerated particles.  Typically, the momentum and energy fluxes are conserved to within 5\% everywhere in the shock structure.  As was shown in, e.g., \citet{EWB2013}, and as will be presented below, these modified shocks result in particle distribution functions that deviate significantly from the traditionally-assumed power law.  The efficient injection we see is nonetheless consistent with PIC results.

In relativistic shocks, the Weibel instability mediates energy transfer from ions to electrons in the upstream region \citep{SSA2013,Ardaneh_etal_2015}.  The effect of this energy transfer is that electrons cross the shock carrying as much as 40\% of the kinetic energy flux.  A basic scheme for approximating this energy transfer was implemented in \citep{WEBL2015}, upon which we iterate here.  Previously kinetic energy was exchanged only at the shock itself, simplifying away that the transfer happens during passage through the Weibel-dominated regions of the precursor.  In this work thermal kinetic energy is subtracted from protons as they pass through the precursor, and added as thermal energy to electrons at the same position.  Although the code only propagates one particle at a time through the shock structure, this implementation mimics simultaneous transfer.

The degree of energy transfer is controlled by a parameter $f_\mathrm{trans}$, which is the approximate fraction of proton kinetic energy that is donated to electrons \citep[called $f_\mathrm{ion}$ in][]{WEBL2015}.  PIC simulations suggest that $f_\mathrm{trans} = 0.4$ for shocks with a Lorentz factor above 10, which value we use for all shocks simulated in this work.  Below a Lorentz factor of 10, the instabilities responsible for energy transfer should be quenched by the higher transverse thermal velocities of the inflowing particles relative to their speed of approach \citep{LemoinePelletier2011}, and so energy transfer should diminish.  Early testing showed that the manner in which $f_\mathrm{trans}$ decreases leaves an observable imprint on the afterglow light curve, and this may offer an important diagnostic connecting observations with the plasma microphysics at the shock.  For now, however, since we know of no PIC simulations or theoretical work describing the turbulent precursor in the mildly relativistic and trans-relativistic regime, we avoid this complication by taking $f_\mathrm{trans}$ to be constant over the entire time covered by our simulations.

To model observations of afterglows and other astrophysical shocks, the Monte Carlo code calculates photon production due to the accelerated population.  Previous work \citep{WEBL2015} included three emission mechanisms, which are used here also: pion decay due to hadron-hadron collisions, inverse Compton radiation, and synchrotron radiation.  However, the implementation of these processes has changed between \citet{WEBL2015} and this work; we provide further details in Section~\ref{sub:photons}.

A significant difference between previous work under this Monte Carlo framework and the results presented here is that we assume that the magnetic field is turbulent and compressed by the shock.  The upstream field is assumed to be turbulent, with components $B_{0\parallel}$ and $B_{0\perp}$.  It is a standard result of magnetohydrodynamics that $B_{\parallel}$ is unchanged across a shock, while $B_{\perp}(x) = B_{0\perp} \cdot z(x)$, where $z(x) = \gamma_{0} \beta_{0} / (\gamma(x) \beta(x))$ is the density ratio between the far upstream and local conditions.  In combination, then, the compressed magnetic field strength within the shock structure is given by
\begin{equation}
  B(x) = B_{0} \sqrt{\frac{1}{3} + \frac{2}{3} z(x)^{2}} ,
  \label{eq:bcomp}
\end{equation}
with $z(x)$ defined as above \citep[e.g.,][]{Reynolds1998}.  As mentioned before, we neglect amplification due to the Weibel instability, and assume that the field has decayed to this value prior to dilution by expansion (see Section~\ref{sub:cooling}).

The Monte Carlo approach as outlined above is both complex and versatile, but it does have several limitations.  We do not make any assumption about the spectrum of magnetic turbulence responsible for our particle scattering, beyond that it exists and causes the prescribed scattering.  At the moment our scattering mean free paths are position-independent, and further do not take into account the expected departure from Bohm diffusion at high energies \citep{PPL2011,PPL2013,EWB2016}.  In the context of GRB afterglows, however, the most severe limitation is the assumption of a steady-state shock.  In the following section we will describe how multiple steady-state simulations are joined to provide a complete picture of an evolving system.

\section{An evolutionary model}
\label{sec:evomodel}

In this section we describe how a steady-state Monte Carlo model is applied to the dynamic environment of a GRB afterglow.  The process has three key parts: (1) the hydrodynamic pre-process, (2) the Monte Carlo simulations, and (3) a post-process that determines both the evolution of downstream material and the overall photon production.

\begin{table*}[t]
\begin{center}
\caption{Afterglow model, common parameters.}
\label{tab:afterglow_common}
\renewcommand{\arraystretch}{1.5} 
\begin{tabular}{lccccccccl}
\tableline \tableline
  Time step\tablenotemark{a}\tablenotetext{1}{For figures that compare data from different time steps against each other, each time step has a color that will be consistently used to represent it.}
&  $t_\mathrm{eng}$\tablenotemark{b}\tablenotetext{2}{Elapsed time in the rest frame of the central engine; also the ISM frame.}
&  $t_\mathrm{obs}$\tablenotemark{c}\tablenotetext{3}{Elapsed time since the start of the GRB for an observer on Earth, assuming a distance of 1~Gpc and cosmological parameters as defined in \citet{Planck2013XVI}.}
&  $R_{\mathrm{FS}}$ 
&  $\gamma_{0}$
&  $\rRH$
&  $N_{g,\mathrm{TP}}$
&  $N_{g,\mathrm{NL}}$
&  $\Pinj$
&  Color in figures\tablenotemark{a}
\\ 

& (sec)
& (sec)
& (cm)
& 
& 
& 
& 
&
& 
\\ 
\tableline
  $\;$1  &  $7.5\xx{6}$  &  $3.0\xx{2}$  &  $2.3\xx{17}$  &  62.8  &  3.001  &  $8\xx{4}$  &  $5\xx{4}$  &  0.06  &  Magenta  \\
  $\;$2  &  $1.0\xx{7}$  &  $9.7\xx{2}$  &  $3.0\xx{17}$  &  40.1  &  3.002  &  $5\xx{4}$  &  $2\xx{4}$  &  0.06  &  Purple  \\
  $\;$3  &  $1.4\xx{7}$  &  $3.2\xx{3}$  &  $4.1\xx{17}$  &  25.6  &  3.003  &  $4\xx{4}$  &  $1.5\xx{4}$  &  0.05  &  Blue  \\
  $\;$4  &  $1.8\xx{7}$  &  $1.1\xx{4}$  &  $5.5\xx{17}$  &  16.4  &  3.006  &  $3\xx{4}$  &  $1\xx{4}$  &  0.05  &  Cyan  \\
  $\;$5  &  $2.5\xx{7}$  &  $3.5\xx{4}$  &  $7.4\xx{17}$  &  10.5  &  3.014  &  $2\xx{4}$  &  8000  &  0.04  &  Green  \\
  $\;$6  &  $3.4\xx{7}$  &  $1.2\xx{5}$  &  $1.0\xx{18}$  &  6.7  &  3.033  &  $1.4\xx{4}$  &  6000  &  0.04  &  Yellow  \\
  $\;$7  &  $4.5\xx{7}$  &  $3.8\xx{5}$  &  $1.4\xx{18}$  &  4.3  &  3.08  &  9000  &  4000  &  0.03  &  Orange  \\
  $\;$8  &  $6.2\xx{7}$  &  $1.3\xx{6}$  &  $1.8\xx{18}$  &  2.7  &  3.18  &  5000  &  2000  &  0.03  &  Red  \\
\tableline
\end{tabular}
\end{center}
\end{table*}

\subsection{The hydrodynamical base}
\label{sub:hydro_base}

The hydrodynamical picture of a GRB jet is akin to the nonrelativistic supernova remnant scenario: a piston of ejecta is driven into the circumburst medium (CBM).  This creates a forward shock propagating ahead of the contact discontinuity (the ejecta--CBM interface) and a reverse shock propagating down through the ejecta.  At present we focus our attention on the forward shock alone; particle acceleration may happen at the reverse shock as it passes through the jet, but we defer this topic to future endeavors.

As long as shock-accelerated particles do not escape the shock in significant quantities, the location of the afterglow's forward shock is mostly independent of acceleration efficiency.  Ambient particles crossing a relativistic shock are heated to relativistic speeds and achieve an adiabatic index $\Gamma_{2} \approx 4/3$.  If particles were able to escape the shock in large quantities, the remaining fluid would be more compressible (lowering the effective adiabatic index) and so the forward shock would stay close to the position of the contact discontinuity \citep{BlondinEllison2001, WarrenBlondin2013, DuffellMacFadyen2014}.  Another possibility is that the shock could radiate away a substantial portion of its kinetic energy.  We will show later that both loss mechanisms are negligible.

It is therefore appropriate to use a purely hydrodynamical model for the location of the forward shock at any given time.  We have chosen to use the Blandford--McKee solution for an adiabatic blast wave \citep{BlandfordMcKee1976}, with slight modifications for the initial acceleration period as described in \citet{KPS1999}.  The location/velocity/age of the analytical Blandford--McKee solution matches very well with numerical results until the forward shock has decelerated to a Lorentz factor of 5, and reasonably well down to the trans-relativistic regime of $\gamma_{0} \approx 2$.  Downstream from the shock the analytical solution diverges from the numerical one because of the approximations made during the derivation; we will discuss these departures later, in Section~\ref{sub:observables}, as they relate to the results presented in Figure~\ref{fig:phot_by_source_NL}.

The Blandford--McKee solution scales with upstream density and isotropic energy, and the modifications proposed in \citet{KPS1999} introduce an additional dependence on the maximum Lorentz factor achieved by the shock.  For the example afterglow we describe here, we use a constant-density ambient medium of one proton per cubic centimeter,\footnote{Despite occurring in the vicinity of a massive star, a significant fraction of GRB afterglows are more compatible with a constant-density CBM than with a wind-like CBM profile \citep[e.g.,][]{PanaitescuKumar2002,Schulze_etal_2011}.} an isotropic-equivalent energy release of $10^{53}$ ergs, and a maximum Lorentz factor of 400.  Where needed, the jet is assumed to have an opening angle of 5$^{\circ}$.  The upstream temperature is taken to be $10^{8}$~K to speed computation.  As long as the Mach number of the shocks remains much greater than unity, the upstream temperature has minimal effects on the Monte Carlo results.  Even at time step 8 the sonic Mach number is approximately 170; all shocks modeled are strong enough to justify our unusually high upstream temperature.  We note that all particles swept up by the shock are nonrelativistic in their local plasma frame, and that our Monte Carlo treatment of shock acceleration does not require particles to be relativistic in order to enter the Fermi process.

Our evolutionary model for GRB afterglows selects various points during the Blandford--McKee solution for detailed modeling with the Monte Carlo code described in Section~\ref{sec:MCcode}.  The specific times, and their associated physical properties such as radius and shock speed, are shown in Table~\ref{tab:afterglow_common}.  The Table also shows select Monte Carlo parameters used to model the shocks.  The only free parameter in Table~\ref{tab:afterglow_common} (i.e. not determined by either the Blandford--McKee solution or basic shock acceleration principles) is $\Ng$.  Convergence testing of the shocks showed that larger values than presented in Table~\ref{tab:afterglow_common} do not significantly affect the results presented in this paper (e.g., doubling $\Ng$ induced a change in the spectral index $p$ of a few hundredths).

At each time step, we model the interaction of particles with a shock whose Lorentz factor is given in Table~\ref{tab:afterglow_common}.  We use two means to limit the maximum particle energy.  First, we enforce a maximum acceleration time that any particle may have, which we set equal to the age of the shock.  Second, we place a free escape boundary (FEB) upstream from the shock.  Any particle that reaches this boundary is allowed to escape the shock in the upstream direction, and is assumed to freely stream away from the shock into the galaxy at large.  We choose 2\% of $R_\mathrm{FS}$ as our FEB location; within this distance the physical curvature of the shock is ignorably small, and our assumption of a parallel plane shock holds.  (In practice, the FEB location is irrelevant.  No particles escape upstream before the final two time steps, and even then the fraction of energy/momentum flux they carry away is negligible.)

\begin{figure}
  \epsscale{1.25}
  \includegraphics[width=\columnwidth]{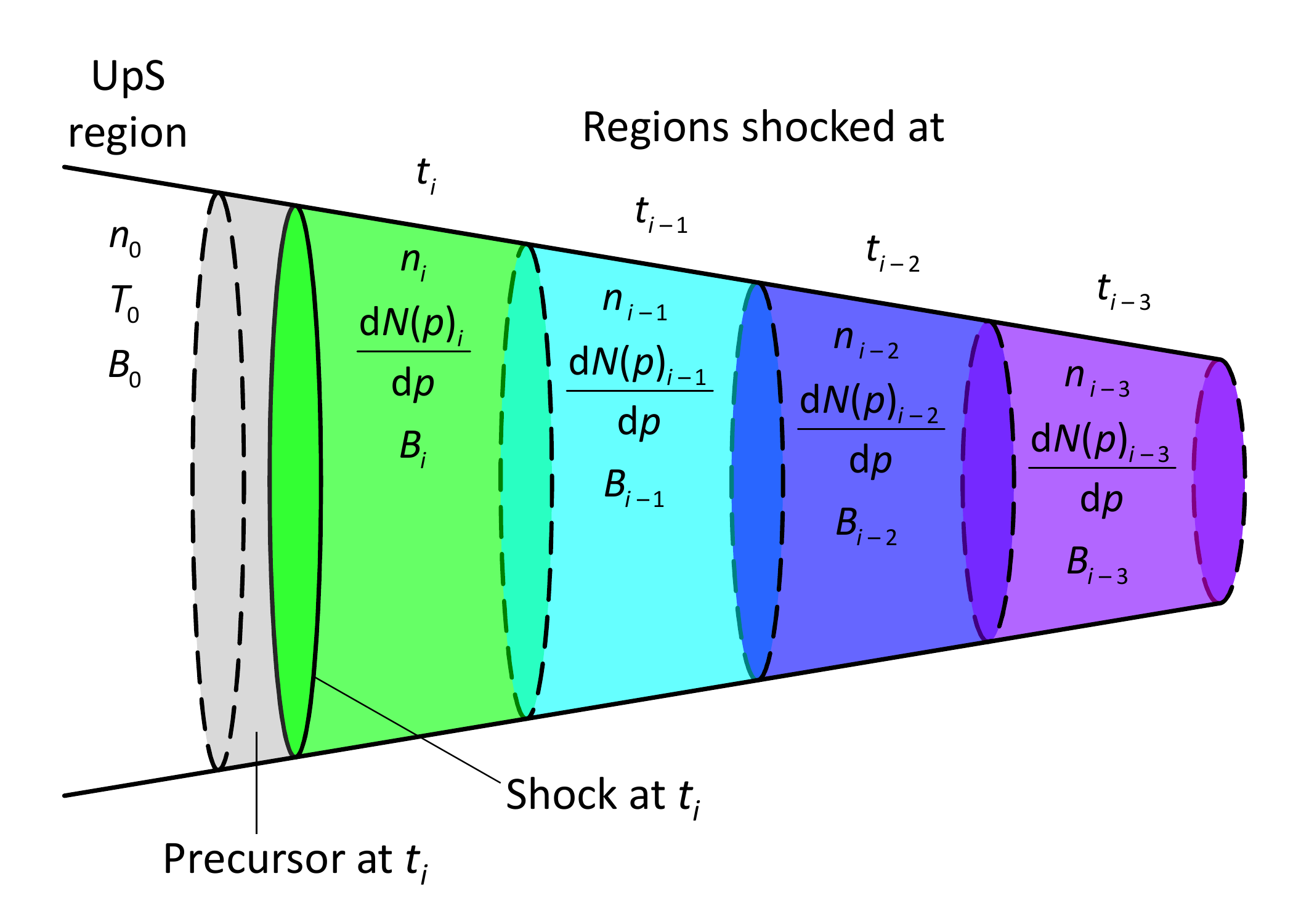}
  \caption{Schematic showing the multi-shell model used in this work.  At each time step, particle spectra and fluid parameters are tracked for both the population of particles currently interacting with the shock (green region and precursor at left of image).  For each shell at each time, we record the density $n$, the particle distribution function $d N/d p$, and the magnetic field strength $B$.  Between time steps all shocked populations are swept downstream from the shock and evolved as described in the text.}
  \label{fig:emission_shells}
\end{figure}

After each time step, shocked plasma is swept downstream as a single shell.  We assume that the fluid in the shell has decoupled from the shock, and that no further acceleration will take place.  The result is illustrated in Figure~\ref{fig:emission_shells}.  At any given time step the shock is interacting with the precursor and the closest downstream region.  Shells that were shocked at previous time steps are arranged in an onion-like structure moving down the jet.  This approach has been used to study the non-relativistic shocks of supernova remnants \citep[e.g.,][]{EDB2004,Slane_etal_2014}.

\subsection{Between time steps}
\label{sub:cooling}

As illustrated by Figure~\ref{fig:emission_shells}, we must keep track of three quantities to model the evolution and downstream emission of previously-shocked particles: the total density, the particle distribution, and the strength of the local magnetic field.

The total density in any particular shell comes from the Blandford--McKee solution.  Both the upstream density and forward shock position are known at all times, so (with the assumption that the jet is conical in shape) it is simple to count the total number of particles swept up at each time step.  We then integrate over the downstream region until the total number of particles matches the total swept up during that time step; the radius at which these two particle counts are equal separates adjacent shells.

The particle distribution function has a fixed normalization, which assumes negligible particle transport between shells.  The shape of $\dNdp$ once the particles decouple from the shock is influenced by two processes: adiabatic cooling and radiative losses.  As the plasma flows away from the shock and expands, on average the particles are scattering off of magnetic turbulence that is receding from them.  The average scattering event therefore results in a loss of energy, and over time a given particle's energy varies as
\begin{equation}
  \frac{dE}{dt} = - (\Gamma - 1) \frac{E}{V} \frac{dV}{dt} ,
  \label{eq:adiab_losses}
\end{equation}
where $E$ is the particle's energy, $\Gamma$ is the adiabatic index of a monoenergetic population of particles at energy $E$, and $V$ is the volume of the shell \citep{Reynolds1998}.  In addition to adiabatic cooling, electrons radiate away energy via synchrotron and inverse Compton losses; despite potentially high magnetic fields in the vicinity of the shock, proton synchrotron is never energetically relevant.

As will be seen later (e.g., Figures~\ref{fig:time_seq_en_frac} \& \ref{fig:phot_by_source_NL}), the highest-energy electrons radiate away a significant fraction of their energy during the interval between time steps.  To accurately track their energies, we divide the first time step after the particle population decouples from the shock into a large number of substeps (only the first time step is treated in this manner; later cooling steps for the same population are handled in one interval).  The duration of each substep is set so that the highest-energy electrons lose no more than $10^{-3}$ of their energy during the substep.  The intermediate electron spectra are stored for photon production, ensuring that we are adequately capturing the rapid cooling of the electrons in the post-shock environment.

Finally, the magnetic field strength is determined by assuming that the total magnetic energy in each shell is constant in time.  The field strength therefore decays as $V^{-1/2}$.  Note that this is a slower decay rate than that experienced by particles in the plasma, so the fraction of energy in magnetic fields increases with time (albeit slowly).

\subsection{Photon production}
\label{sub:photons}

As mentioned in Section~\ref{sec:MCcode}, we use three processes to determine photon emission: synchrotron, inverse Compton, and pion decay due to proton-proton collisions.  These three processes happen in different reference frames, however, and additional relativistic considerations affect the flux observed at Earth.  We elaborate here, and note that each emission shell is handled independently at each time step.  Additionally, we assume that all emitting populations are isotropic in their local plasma frame; this will not be the case very near the shock, but is true for a large majority of the particles being considered.

Both synchrotron and pion decay occur in the fluid rest frame.  For synchrotron radiation \citep{RybickiLightman1979}, we assume that both the electron distribution and the magnetic field turbulence are isotropic.  This clearly is not the case upstream from the shock (Weibel turbulence creates an ordered field, and cosmic rays necessarily have a highly anisotropic distribution in order to have scattered ahead of a relativistic shock front); for the downstream regions under consideration presently, though, this is a reasonable assumption.  The kernel used to compute the pion production cross section in $p-p$ collisions \citep{Kafexhiu_etal_2014} assumes that target particles are at rest in the local plasma frame; for an assumed isotropic distribution of cosmic ray protons this is true on average.  The photon distributions produced by these two processes may be transformed into the engine (i.e., ISM) frame using standard special relativistic equations.

The case of inverse Compton emission is somewhat more involved, as multiple photon fields are present: the local synchrotron photons, the cosmic microwave background (CMB), and the interstellar radiation field (ISRF).  Synchotron self-Compton (SSC) emission happens in the local plasma frame and is transformed alongside synchrotron and pion decay emission.  The photon field used for SSC is the synchrotron field of the local emission shell: at a given location the number density of locally-produced synchrotron photons is much greater than the density of those produced in adjoining shells, so we ignore photon transport between shells.

Inverse Compton using the CMB or ISRF takes place in the engine frame, so the electron distributions must be transformed into that frame before photon production can be computed.  The CMB is a thermal photon field at a temperature of 3.35~K and energy density $\approx 0.60$~eV \pcc\ (appropriate for a GRB at redshift $z \approx 0.23$).  The ISRF is highly dependent on the environment of the progenitor star, but the work presented here uses the $R = 8$~kpc ISRF presented in Figure 1a of \citet{Porter_etal_2008}.  This field is appropriate for a relatively isolated progenitor star (i.e., not part of a compact cluster) in a Milky Way-like host galaxy.  As will be shown later, however, the IC-ISRF luminosity is many orders of magnitude lower than that due to SSC emission.  The exact shape and normalization of the ISRF is unlikely to be observationally important.

All of the preceding calculations are performed to arrive at a luminosity for the entire emitting shell.  Additional calculations are necessary to convert this luminosity into a flux as observed at Earth.  When shifting from the plasma frame to the engine frame, relativistic aberration beams half of the emission into a cone whose half-opening angle is given by $\cos(\theta_{h}) = \beta_\mathrm{rel}$, where $\beta_\mathrm{rel}$ is the relative velocity between the two frames (in the limit of $\beta_\mathrm{rel} \rightarrow 1$, this formula reduces to the more traditional $\theta_{h} \sim 1/\gamma_\mathrm{rel}$).  This is also the half-opening angle for an observer looking at the jet, so (at early times) an observer does not see emission from the entire jet surface.  We compute the fraction of the jet that lies within the angle $\theta_{h}$ and modify the total luminosity accordingly.  We then compute the area of the $\theta_{h}$ cone at the appropriate luminosity distance (1 Gpc) to convert from luminosity to observed flux.

Our photon spectra are additionally processed to model absorption at high energies, but not at low energies.  Since the simulated afterglow occurs at cosmological distances, pair production off of the extragalactic background light (EBL) absorbs energy flux at GeV or higher energies.  We use the opacity tables presented in \citet{FRV2008} to calculate this absorption.  At the low-energy end of the spectrum, synchrotron self-absorption (SSA) is the dominant absorption mechanism.  This process occurs locally within the shocked material, where the velocity difference from back to front may be great enough to induce significant anisotropies in the absorbing electron population.  We do not model this more complicated absorption process here, but neither will we draw conclusions from the radio to infrared parts of our photon spectra.

\section{Example afterglows}
\label{sec:afterglows}

Having explained the Monte Carlo code (Section~\ref{sec:MCcode}) and the manner in which it is coupled to a hydrodynamical simulation to create an evolving afterglow model (Section~\ref{sec:evomodel}), we will now present the results from three test cases.  The first two models assume that acceleration takes place in the test-particle limit, with few enough particles accelerated that the structure of the shock is not significantly modified.  The particular values of $\Pinj$ are listed in Table~\ref{tab:afterglow_common}.  In one model, thermal (uninjected) particles are explicitly excluded from the radiating population, though they contribute to the pressure and energy density downstream.  We refer to this as the ``CR-only'' test case.  In the second model, the ``TP'' model, all particles are included in the photon production calculations.  In the third model, ``NL,'' injection is efficient enough that the nonlinear backreaction of accelerated CRs on the shock structure cannot be ignored.

We must comment here on the physical reasonableness of all three models. The CR-only case is \textit{not} intended to be physically plausible.  There is no way to conserve number, momentum, and energy fluxes in a shock interacting with the ISM unless a thermal population is present.  The CR-only case is included only because it mimics the standard synchrotron model for afterglow emission, which ignores the thermal population.  It serves as a control against which our other two models may be compared; we are not presenting it as a valid alternative to either the TP or the NL model.  Both the TP and NL models are physically possible, and operate at different ends of the injection efficiency spectrum.

All three model afterglows use an upstream ambient magnetic field $B_{0} = 3$~mG.  This is orders of magnitude higher than the typical value of 3~\muG\ assumed for the Milky Way.  There are locations where such high fields may exist in the circumburst environment,\footnote{Specifically, \citet{RQH2008} report magnetic field strengths upwards of 10~mG in starburst galaxies from measurements of Zeeman splitting of the 1667 MHz OH megamaser line.  Additionally, ambient field strengths within 100-200 pc of the center of the Milky Way may locally reach or exceed mG levels \citep{Ferriere2009}.} but this value was chosen for a different reason: once it is compressed according to Equation~\ref{eq:bcomp}, the downstream field approaches an energy fraction of $\epsB \approx 10^{-3}$, when both rest mass energy and pressure are considered.  This is lower than the traditional value, $\epsB \sim 10^{-2}$, but in line with current expectations \citep[it is potentially even on the high side; see][and see also the discussion of $\epsilon_{B-}$ in \citep{LLW2013}]{SBDK2014,Beniamini_etal_2015}.  Because we assume only magnetic field compression and no additional amplification, the value of $\epsB$ just downstream of the shock (that is, prior to the expansion described previously) decays with time as the shock decelerates---see Equation~\ref{eq:bcomp}---rather than maintaining a constant value at all times.  This temporal decay at the shock front occurs in addition to the expected dilution of magnetic energy density as the plasma expands downstream from the shock.

\subsection{Particle spectra and energy densities}
\label{sub:unobservables}

\begin{figure}
  \epsscale{1.15}
  \includegraphics[width=\columnwidth]{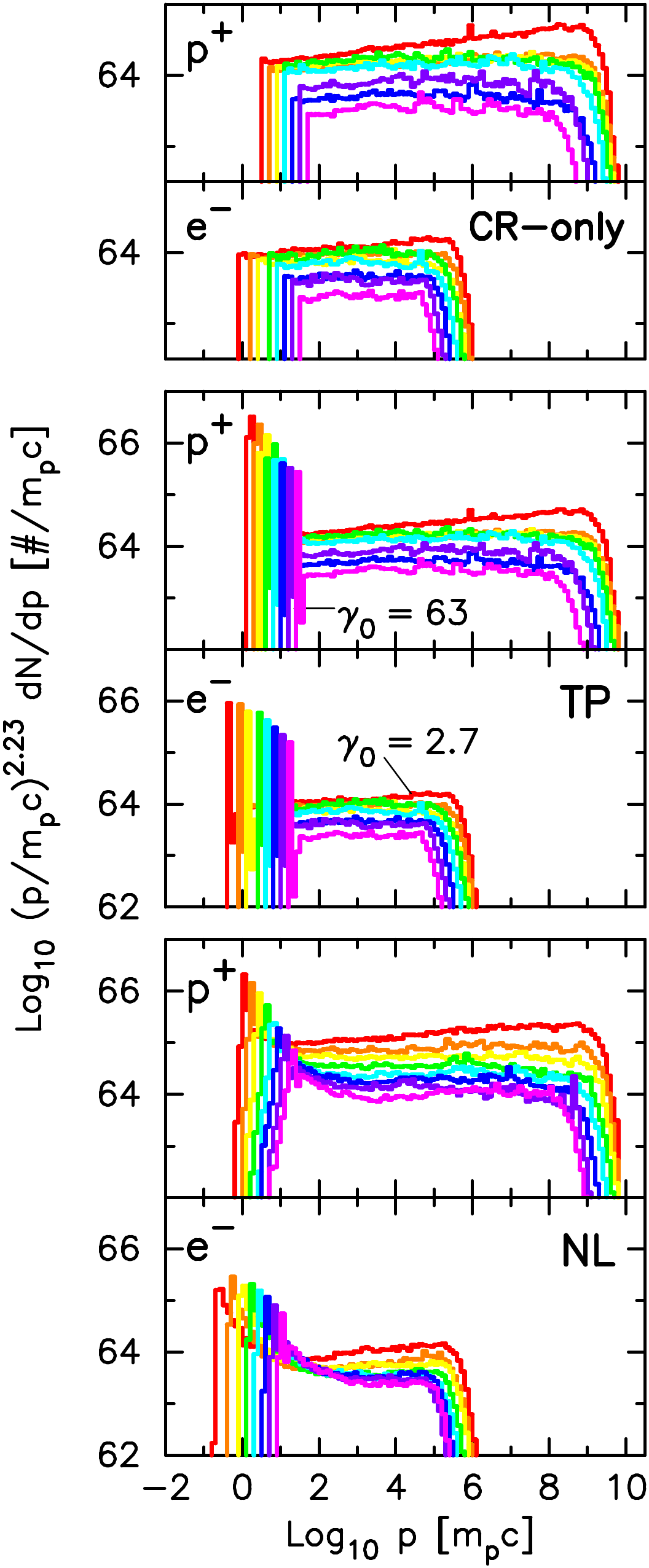}
  \caption{Particle spectra for each time step, prior to any cooling, for the CR-only case (top two panels), TP model (middle two panels), and NL model (bottom two panels).  In each pair of panels, the top panel shows the proton spectra and the bottom panel shows the electron spectra.  All momenta are in the local plasma frame, and colors refer to time steps as described in Table~\ref{tab:afterglow_common}.}
  \label{fig:spectrum_compare}
\end{figure}

In Figure~\ref{fig:spectrum_compare} we show the spectra of cosmic rays for all three models, in the local plasma frame.  For the sake of clarity, the cooled populations that have been swept further downstream (the interior shells of Figure~\ref{fig:emission_shells}) are not given.  These spectra were taken just downstream from the subshock at $x = 0$, to most closely match the spectral indices predicted by analytical work.  Though we assume an isotropic distribution of particles for all photon production processes, this is obviously not the case at the subshock.  All particles require several diffusion lengths of transport downstream before their angular distribution relaxes into isotropy.  Since the diffusion length is an increasing function of momentum (Equation~\ref{eq:lambdamfp}) higher-energy particles remain anisotropic for longer distances downstream.  The effects of anisotropic particle distributions will be addressed in future work; for now we merely note the assumption being made.

The spectra presented in Figure~\ref{fig:spectrum_compare} do not use arbitrary units.  Since upstream escape is negligible (no particles escaped during the first six time steps, and even in the last two steps the escaping flux was less than $10^{-6}$ of the incoming energy flux), and since the difficulty of scattering upstream against a relativistic shock means most particles are downstream, the total number of particles in each spectrum is the number of particles swept up at a given time step.  As such, the spectra are not only absolutely normalized within each model: they are normalized between models as well.

Both the CR-only case and TP model result in very simple spectra.  This is unsurprising, as they were designed to reproduce the simple power laws used in the standard synchrotron model for afterglows.  In the TP model's spectra, there is a thermal peak\footnote{\label{ftnt:thermal_def}The term ``thermal'' is used only to specify those shocked particles that don't enter the Fermi acceleration process.  The results we obtain in no way require or presume that these particles equilibrate in a thermodynamic sense.} of particles that crossed the shock once (but did not enter the acceleration process), and a high energy tail of shock-accelerated particles out to a maximum energy.  The thermal peak has been excised from the CR-only case, while keeping the normalization of accelerated particles the same.  In all cases the spectral index of the proton distributions decreases with time, as the shock Lorentz factor drops \cite[see][]{KeshetWaxman2005,Ellison2005}.

Since protons experience no losses while they interact with the shock, the high-energy turnover of the CR tail is due to the limit on acceleration time.  The electrons, by contrast, experience significant synchrotron losses in the intense magnetic fields downstream of the shock.  The slow increase in the maximum energy of the electron spectrum is due to decreased compression also lowering the magnetic field and cooling rate.

Our Monte Carlo model predicts that protons are easily accelerated beyond the knee in the cosmic ray spectrum; the high magnetic fields, combined with pitch-angle scattering, confine the protons to a small volume around the shock, and allow for many shock crossing cycles in the time-limited scenario presented here.  The proton spectrum extends above $10^{17}$~eV even at the earliest time step shown in Figure~\ref{fig:spectrum_compare}, and the maximum energy slowly increases in successive time steps.  This is a consequence of our magnetic field structure, which is uniformly strong in the entire scattering region downstream from the shock.  Both analytical studies and PIC simulations predict that the magnetic field strength should decay with distance from the shock; this will increase the scattering length of particles, allow them to spend more time far from the shock, reduce the number of shock crossing cycles they can complete, and so lower the maximum energy.  A more detailed study of this model's predictions for ultra-high energy cosmic rays, with a more realistic magnetic field structure, will be published elsewhere.

Figure~\ref{fig:spectrum_compare} also shows the particle spectra produced by the NL shocks.  While much of the preceding discussion (regarding, e.g., spectral indices and maximum energies) applies to the NL model as well, there are a few important points of comparison.

First, the thermal peaks of the two sets of spectra occur at roughly the same momenta for a given time step.  The NL model's peaks are always at slightly lower momentum than the TP model's peaks; this is a well-known result that aids in reducing the downstream fluxes and ensuring conservation.  At the high-energy end, the TP and NL models are almost identical.  Both models' electron spectra turn over at the same energy (to within one or two bins, a factor of 1.6 or less), despite the spectra extending over 4-7 decades.

The change from unmodified to NL shocks affects the shape of the particle spectra.  Where the unmodified shocks generate simple power laws above the thermal peak, the velocity gradients in the NL shocks induce more structure.  The thermal peaks are still present (albeit at slightly lower energies thanks to momentum and energy conservation), but there is now a substantial steepening of each spectrum before it levels out into something approaching a power law.  Above a certain energy all particles have long enough mean free paths that they can effectively ignore the velocity gradient of the shock precursor during their scattering.  At this point all both models' spectra settle into power laws with identical spectral indices (albeit different normalizations relative to the thermal peak).

\begin{figure}
  \epsscale{1.15}
  \includegraphics[width=\columnwidth]{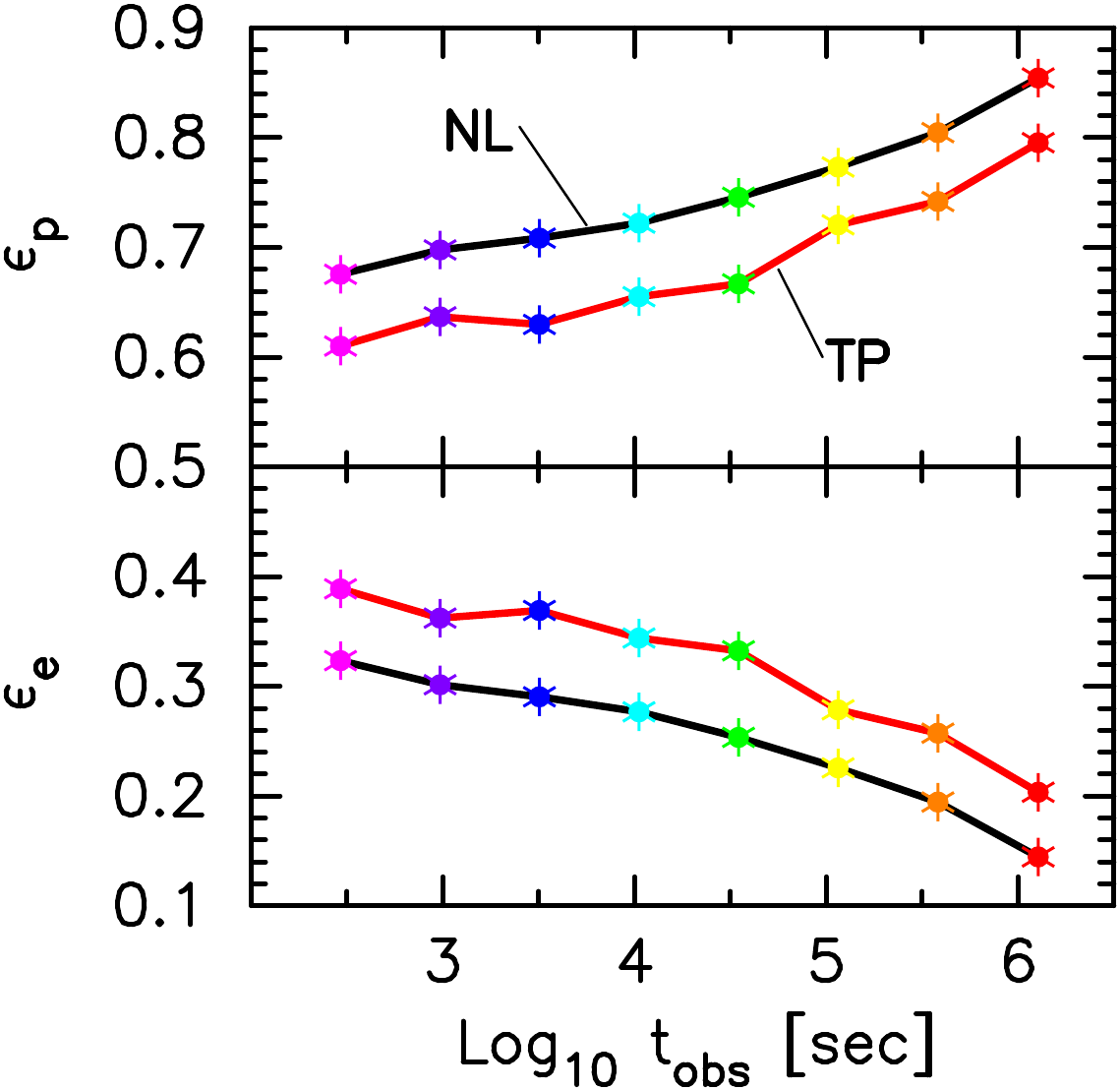}
  \caption{Fractional energy density in the outermost shell as a function of time.  The TP model is marked with a red line, and the NL model with a black line.  The colored points mark the values at time steps described in Table~\ref{tab:afterglow_common}.  \textit{Top panel}: $\epsilon_{p}$, the fractional energy density in protons.  \textit{Bottom panel}: $\epse$, the fractional energy density in electrons.}
  \label{fig:time_seq_en_frac}
\end{figure}

Both particle spectra and magnetic field strength are known everywhere in the jet at all times.  This allows us to calculate directly, rather than infer from observations, two important parameters in the standard synchrotron model: $\epse$ and $\epsB$, the fractional energy density in electrons and the magnetic field, respectively.  Additionally, we calculate the time dependence these quantities have, rather than fixing them to be constant throughout the afterglow.  Specifically, we define
\begin{equation}
  \epsilon_{s} \equiv \frac{E_{s}(i,t)}{E_{e}(i,t) + E_{p}(i,t)} ,
  \label{eq:epse}
\end{equation}
where $s$ could be either $e$ (for electrons) or $p$ (for protons), and
\begin{equation}
  E_{e}(i,t) = \int \frac{dN_{e}(p,i,t)}{dp} E(p) dp
  \label{eq:Eit}
\end{equation}
is the total energy (i.e. $E(p) = m_{e}c^{2}\sqrt{1 + (p/m_{e}c)^2}$), of all electrons (both thermal and accelerated) in cohort $i$ at time $t$.  The value $\epsilon_{p}$ is defined analogously, with $E(p) = m_{p}c^{2}\sqrt{1 + (p/m_{p}c)^2}$.

The above definition of $\epse$ is an extension in two important respects of the value commonly used in the GRB literature.  First, values of $\epse$ reported in the literature are assumed to be the fractional energy density in electrons just after crossing the shock.  In the one-zone approach used in the standard 
synchrotron model, the electron energy density is fixed to this value.  In our multi-zone approach we follow the downstream evolution of $\epse$.  The second point of interest about Equations~\ref{eq:epse} and \ref{eq:Eit} is that they explicitly include rest mass-energy.  In the standard synchrotron model all particles are ultra-relativistic, so contributions of rest mass-energy to the energy density are minimal.  As the shock slows, the rest mass-energy of the protons can no longer be ignored compared to their kinetic energy, and a decision must be made whether to include it.  We choose to include rest mass-energy in our calculation of energy density, and note that $\epsB$ is also defined relative to the total, not simply the kinetic, energy density.  The inclusion of rest-mass energy is largely irrelevant at early times, when all particles are highly relativistic.  At later times, when the thermal particles have plasma-frame Lorentz factors of just a few, including rest-mass energy enhances the contribution from protons.

The values of $\epse$ and $\epsilon_{p}$ are plotted in Figure~\ref{fig:time_seq_en_frac} for the TP and NL models, for the outermost shell, noting again that the simulation follows their evolution throughout the simulation.  We do not distinguish between cosmic-ray and thermal particles for either model here.  (The distinction between these two populations is obvious in the case of the TP model.  As shown in Figure~\ref{fig:spectrum_compare}, the NL model's thermal peaks are smoothly joined to the accelerated population, making it unphysical to sharply distinguish where the former group ends and the latter begins.\footnote{The Monte Carlo code, of course, allows us to separate CR and thermal particles based on whether or not they have completed at least one up--down--up cycle.  While this is useful for tracking injection rates, it is unlikely to be physically meaningful given how smoothly connected the thermal and accelerated populations are.})  The CR-only case, where $\epse$ is nearly constant at $\sim 0.4$ for the entire time modeled, is not drawn in Figure~\ref{fig:time_seq_en_frac}.

The difference between the TP and NL models is a few percent, but the TP model consistently places a larger fraction of the bulk kinetic energy into electrons than does the NL model.  This effect occurs in spite of the fact that the electron-to-proton injection ratio is higher for the NL model than for the TP model.  Finally, we note that the effect of including the rest mass energy in Equation~\ref{eq:Eit} is plainly visible in Figure~\ref{fig:time_seq_en_frac}: the value of $\epse$ drops from nearly 0.4 at the start of the afterglow to roughly 0.2 at the end.

\subsection{Photon spectra}
\label{sub:observables}

\begin{figure}
  \epsscale{1.15}
  \includegraphics[width=\columnwidth]{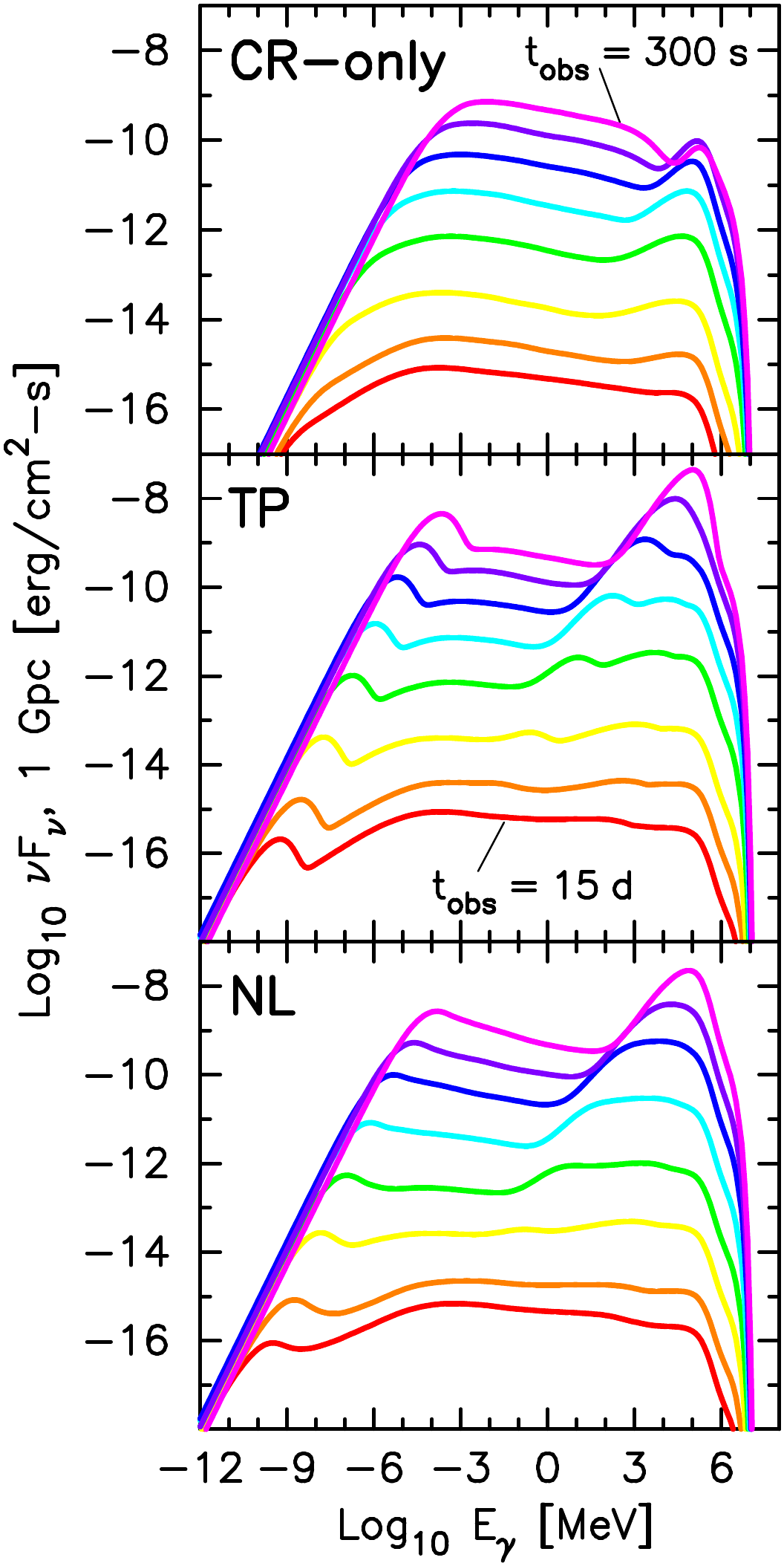}
  \caption{Photon spectra for all time steps and all models.  Colors refer to time steps as described in Table~\ref{tab:afterglow_common}.  Absorption due to the extragalactic background light has been included at high energies, but synchrotron-self absorption at low energies has not.}
  \label{fig:phot_compare}
\end{figure}

In Figure~\ref{fig:phot_compare} we show photon spectra, all scaled to a source distance of 1~Gpc, produced by the CR-only case and the TP \& NL models.  The spectral shapes and normalizations vary substantially between models and with time for any given model.  In general, at low energies there is a steep rise (but see earlier warnings in Section~\ref{sub:photons} about the lack of synchrotron self-absorption).  The synchrotron peak, due to either thermal particles or the low-energy end of the CR spectrum, lies just beyond that, followed by the synchrotron spectrum of the non-thermal particles.  There is a break in the spectra where synchrotron emission ceases (10-100~MeV) and synchrotron self-Compton begins (strongly time-dependent), and finally a turnover due to EBL absorption at energies above 100 GeV.

One obvious difference between the TP/NL spectra and the CR-only spectra is the contribution of thermal particles to the synchrotron spectrum.  Where the CR-only spectra show a smoothly joined broken power law, the TP and NL spectra show a great deal of structure at the low-energy end.  As suggested by Figure~\ref{fig:spectrum_compare}, the extreme drop in normalization between the thermal and nonthermal populations of the TP model causes a factor of 10 difference between the synchrotron thermal peak and the portion of the spectrum due to the nonthermal tail.  The bridge present in the NL model's electron spectra appears in Figure~\ref{fig:phot_compare} as well.  While there is still a thermal peak and a slight steepening of the photon spectrum above it, the transition is much smoother for the NL spectra than for the TP spectra.

In both the TP and NL models, the location of the thermal peak shifts towards lower energies with time as the electrons encounter both progressively weaker magnetic fields and a smaller initial energy boost from crossing slower shocks.  Though our afterglow models begin at an observer time of $\approx 300$~s, it seems reasonable that at earlier times the thermal peak would be at higher energies.  (The same is true for the CR-only case, with a replacement of ``thermal peak'' by ``minimum Lorentz factor.'')  We therefore suggest that a plateau or brightening in emission in a particular waveband might be due to the passage of an unabsorbed thermal peak, especially if it happens prior to 1000~s.\footnote{This is similar, but not equivalent, to the commonly-discussed passage of $\nu_{m}$---the characteristic synchrotron frequency of electrons at the base of the power law distribution---through a particular energy range.  One difference is that the thermal peak is broader than the Lorentz factor $\gamma_{m}$ associated with $\nu_{m}$.  Additionally, the shape of the particle spectra above the thermal peak differs from the power-law case, showing a steeper decay before flattening out into the expected spectral index.  See additional discussion surrounding Figure~\ref{fig:color_compare}.}  This is, of course, within the same time period that emission from a reverse shock might be expected \citep[e.g.][]{SariPiran1999ApJ520,Japelj_etal_2014}, and extending our simulations to earlier times should allow us to characterize the rise time of the emission---assuming, of course, that these energies are not absorbed through the SSA process.

Another clear difference between the models with and without thermal particles is the quantity of photon production at high energies.  When thermal particles are allowed to participate in emission processes, the change in SSC flux is significant.  At time step 1 ($t_\mathrm{obs} = 300$~s), the TP and NL models produce nearly 1000 times more flux at 100~GeV than does the CR-only case.  The peak is high enough for the TP and NL models that the SSC contribution to the photon spectrum extends well below the turnover of the synchrotron spectrum at energies of 10-100~MeV.  The SSC thermal peak is bright enough that it is still detectable over the synchrotron emission days after the GRB, when the peak has dropped to almost X-ray energies: see the curves in Figure~\ref{fig:phot_compare} for time steps 5 \& 6 (green and yellow).

The high-energy part of the TP and NL models' spectra simultaneously explains two open questions regarding $\gamma$-ray emission from GRBs and their afterglows.  First, the peak of the SSC spectrum is roughly 100~GeV at $t_\mathrm{obs} = 300$~s, very close in timing and energy to the highest-energy photon ever detected in association with a GRB, the 94~GeV photon observed 240~s after the trigger of GRB 130427A \citep{Fan_etal_2013}.  That the peak is due to thermal electrons scattering thermal synchrotron photons leads to the second consequence.  The highest energy photons should arrive very soon after, but not simultaneously with, the trigger, since the forward shock takes time to develop and accelerate to its peak Lorentz factor \citep{SPN1998}.  This delay has also been reported by \citet{Ackermann_etal_2013} for a larger sample of GeV-bright GRBs.

\begin{figure}
  \epsscale{1.15}
  \includegraphics[width=\columnwidth]{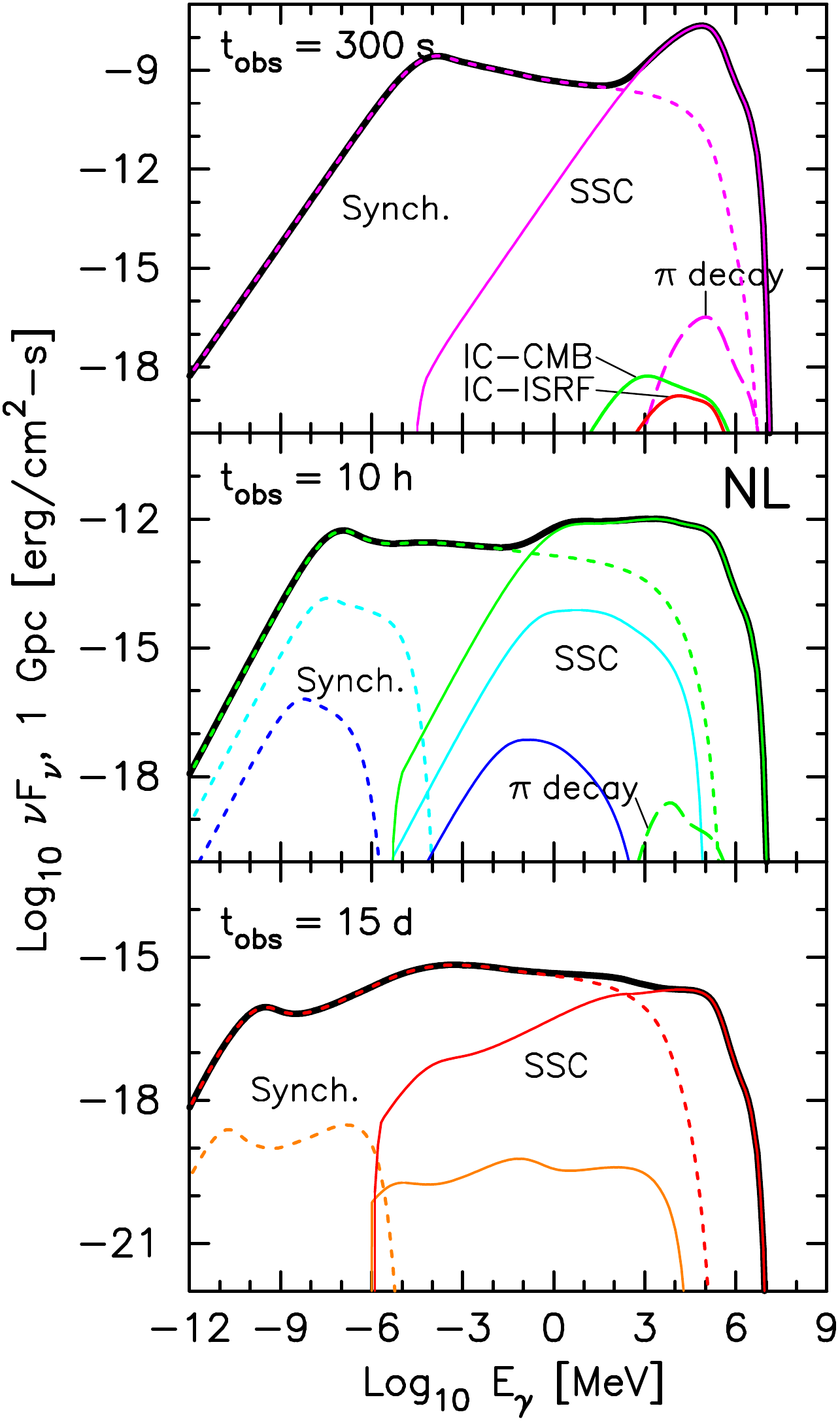}
  \caption{Observed flux due to each shell of shocked particles for selected time steps of the NL afterglow.  With the exception of the IC-CMB and IC-ISRF curves in the top panel, colors refer to emission from cohorts shocked at individual time steps, as described in Table~\ref{tab:afterglow_common}.  The thick black line is the total flux for that time step, as presented in Figure~\ref{fig:phot_compare}.  The dotted lines show emission due to the synchrotron process, thin solid lines show SSC emission, and the dashed line is pion decay.  \textit{Top panel}: time step 1, at $t_\mathrm{obs} = 300$~s.  The green and red curves at the bottom show IC emission off of the CMB and ISRF, respectively.  \textit{Middle panel}: time step 5, at $t_\mathrm{obs} = 10$~h.  The IC-CMB and IC-ISRF curves lie below the shown area.  \textit{Bottom panel}: time step 8, at $t_\mathrm{obs} = 15$~d.  At this time, the pion decay flux is also too faint to be plotted.}
  \label{fig:phot_by_source_NL}
\end{figure}

We plot in Figure~\ref{fig:phot_by_source_NL} the contribution of each shocked cohort of particles for time steps 1, 5, and 8.  Although we did not make the assumption of a one-zone model for emission, it is clear that cooled shells do not contribute significantly to the observed flux.  This conclusion is, of course, dependent on the size of our emission shells.  Using shells that are narrower in time (either due to higher downstream densities or to shorter time steps) would necessarily increase the prominence of earlier shells for a given time step.

The hard $\gamma$-ray regime is dominated by SSC emission in all three models.  (However, the photon flux is low enough that detections are unlikely after the first few hours, consistent with existing non-detections.  See the following section.)  Both hadronic processes and inverse Compton emission off of the CMB/ISRF are present, but many orders of magnitude weaker than SSC.  However, the curves in Figure~\ref{fig:phot_by_source_NL} show that multiple cohorts of particles may contribute to the observed flux, unlike the X-ray afterglow and similar to the optical case.  This is due to the large population of thermal particles, which does not significantly cool during its post-shock evolution.

Examination of Figure~\ref{fig:phot_by_source_NL} shows that the SSC process may be an important mechanism for cooling particles downstream.  Initially (the first time step after a cohort of electrons decouples from the shock) synchrotron losses dominate.  At the next time step after decoupling, SSC grows in importance, with the SSC luminosity approaching the synchrotron luminosity for early cohorts.  The ratio of SSC to synchrotron luminosity falls monotonically after that.  That this ratio approaches unity at any point suggests that a truly self-consistent simulation would include SSC as a cooling process during and after acceleration.  Incorporating SSC as a cooling mechanism is a decidedly nontrivial procedure, however, as the electrons will cool off of their own synchrotron emission; a large fraction of the electron-photon interactions may occur in the Klein-Nishina limit, complicating matters further.  The interaction is nonlinear and challenging to compute even in the case of a single impulsive energization \citep{Schlickeiser2009}, let alone as a continuous process throughout the electrons' acceleration.  We defer this thorny extension to future work.

We note that in all models presented here, the total energy radiated away by the particles over the duration of the simulated period is at most a few percent of the total kinetic energy of the shock.  Our assumption of an adiabatic blast wave \citep[and use of that particular solution from][]{BlandfordMcKee1976} is \textit{a posteriori} justified.

\subsection{Comparison against observations}
\label{sub:discussion}

In this section we will compare our models against observations of afterglows at several wavelengths, and discuss implications for the larger picture of afterglow emission.  Discussions of light curves and spectra follow the convention $F_{\nu} \propto t^{-\alpha} \nu^{-\beta}$, where the temporal and spectral indices $\alpha$ and $\beta$ characterize the observed flux density throughout the afterglow.  We emphasize that we are \textit{not} attempting to fit any particular afterglow here.  Instead, we focus on predicting general evolutionary trends, leaving detailed fits of particular GRBs for future work.

\begin{figure}
  \epsscale{1.15}
  \includegraphics[width=\columnwidth]{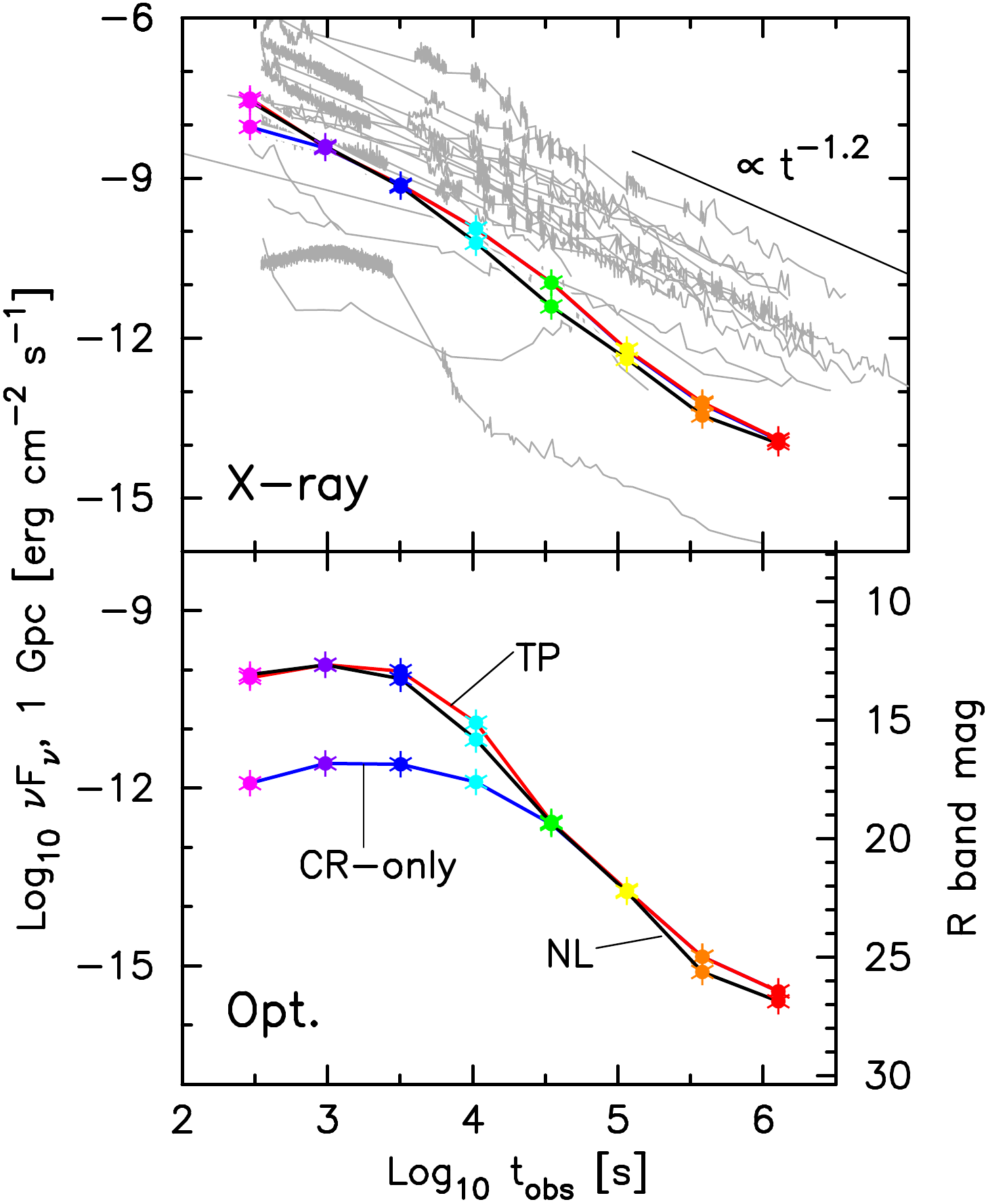}
  \caption{\textit{Top panel}: energy fluxes of the two model afterglows in the \textit{Swift} XRT energy range of $0.3 - 10$~keV.  Also plotted are selected (see text) GRB light curve data from the \textit{Swift} XRT GRB light curve repository.  The traditional $\alpha = 1.2$ decay is illustrated at top right.  The light curve of the CR-only case is marked with a blue line, that of the TP model with a red line, and the light curve of the NL model with a black line; the colored points are fluxes at time steps as described in Table~\ref{tab:afterglow_common}.  \textit{Bottom panel}: optical light curves for the \textit{r} band.}
  \label{fig:optxray_compare}
\end{figure}

In the top panel Figure~\ref{fig:optxray_compare} we plot the total energy flux for the three models in the $0.3-10$~keV band---that is, the \textit{Swift} XRT range---as a function of time.  For comparison we also plot the X-ray light curves of a small (and not exhaustive) collection of GRBs whose afterglows were detected in the radio, optical, and X-ray.\footnote{The GRBs plotted are 050525A, 050724, 050730, 050820A, 051109A, 051111, 060218, 061121, 071122, 080319B, 080810, 081007, 081203B, 090418A, 090424, 090618, 090715B, 091020, and 130427A.  Claims of detection relied on the \textit{Swift} GRB table (http://swift.gsfc.nasa.gov/archive/grb\_table.html/).  In the interest of clarity, we did not plot every burst with XRT/UVOT/radio data.  Some 30 additional bursts that occurred after GRB 091020 were excluded, with GRB 130427A included because of its exceptional nature.}  The observed X-ray light curves have scaled to a distance of 1 Gpc, in keeping with the distance of our simulated afterglows (for GRB 081203B, where no redshift was available, we assumed $z = 2$).

It is clear that the three models are broadly consistent with the observed X-ray afterglow sample, though they are on the fainter side of the distribution.  The TP and NL models track each other very closely, a consequence of the previously-discussed result that the high-energy peaks of both models' electron distributions are very close in both energy and number, throughout the time period considered.  All three models decay faster than $t^{-1.2}$; some of this is due to our assumption that $\epsB$ is non-constant, since $F_{\nu} \propto \epsB^{1/2}$ and our $\epsB$ drops with time.

At the start of our simulated afterglows, the synchrotron emission is being produced by electrons in the thermal peak of the particle distribution.  Since the CR-only case does not have these thermal electrons, its X-ray emission is suppressed compared to the TP and NL models.  After that the X-ray flux is produced by the non-thermal particles, and so the X-ray light curve reflects the structure of the particle distribution producing it.  The TP and CR-only light curves overlap in the top panel Figure~\ref{fig:optxray_compare}, and show a steepening from $\alpha = 1.5$ to $\alpha = 2.1$ around $t_\mathrm{obs} \approx 10^{4}$ seconds, reflecting the passage of a cooling break in the photon spectrum.  This break is obscured by the non-thermal bridge in the NL model, which shows a smooth decay of $\alpha = 1.9$ until $t_\mathrm{obs} = 10^{5.5}$~s.

The optical light curves of the three afterglows are given in the bottom panel of Figure~\ref{fig:optxray_compare}.  It is again the case that the CR-only case behaves very differently from either of the two models that include a thermal population.  At early times emission is suppressed because of the lack of thermal particles contributing in the optical band (see Figure~\ref{fig:phot_compare}), by a factor of roughly 100.  This leads to different temporal behavior for the models based on the presence or absence of a thermal population.

Assuming optical emission isn't blocked by SSA, the CR-only case shows rising behavior in the first hour.  The temporal index $\alpha = -0.46$, close to the theoretical value of $-0.5$ predicted in \citet{SPN1998} and identical to that presented in \citet{GianniosSpitkovsky2009}, who also included a thermal population.  After the rise is a broad (factor of 10 or so in time) plateau associated with the passage of the minimum Lorentz factor.  In both the TP and NL models, this plateau is already underway at $t_\mathrm{obs} = 300$~s, and should occur over a longer time due to the broadness of the thermal peak relative to the sharp rise of the CR-only particle spectra.

After the passage of the thermal peak (or minimum Lorentz factor), optical emission is due solely to the particles in the nonthermal tail of the spectra.  During this time, the CR-only case has a temporal decay index of $\alpha = 2.2$.  Given that the spectral index of electrons is $p = 2.2$, the scaling $\alpha = 3(p-1)/4$ \citep{GranotSari2002} would suggest $\alpha = 0.9$.  The temporal behavior of the TP and NL models is even more divergent from the standard synchrotron model.  The temporal decay index for the TP model is as steep as $\alpha = 3.3$ before flattening out to match the CR-only case.  This steepening is caused by the large difference in normalization, as well as the sharp transition, between the thermal and nonthermal populations in the TP model.  In the NL model where the particle spectra are smoother, the decay is slower, with $\alpha = 2.4$ from time step 3 to time step 7 (approximately two decades in time)

Both the TP and NL models' optical ($r$ band) light curves peak at 12th magnitude, or about 50 mJy (5 mJy if moved to a redshift $z = 2$).  If compared against the normalized optical afterglows presented in \citet{PanaitescuVestrand2011}, the optical light curves would fall on the dividing line between curves with a peaked shape and curves with a plateau.  They are on the low end of the modeled peak brightnesses presented in \citet{Japelj_etal_2014}, especially after accounting for a normalized distance to the bursts.  As these bursts were specifically selected for the presence of a reverse shock component, our results echo their conclusion that forward shock emission alone is unlikely to explain the bright, early optical peaks such as those observed in GRBs 990123 \citep{SariPiran1999ApJ517L}, 061126 \citep{Gomboc_etal_2008}, and 080319B \citep{Bloom_etal_2009}.

We reiterate here that any sort of geometrical or hydrodynamical effects that might cause a jet break are not included in our model; the break in the optical light curve arises naturally out of the acceleration process and associated post-shock evolution.  The wide variety in optical afterglows for nearly identical X-ray afterglows could also explain so-called ``uncoupled'' afterglows \citep{PanaitescuVestrand2011}.  As the behavior of these light curves is tightly linked to the efficiency of injection into the acceleration process, varying this efficiency could also explain the lack of a canonical optical afterglow in comparison to the traditional X-ray afterglow \citep[discussed in, e.g.,][]{Nousek_etal_2006,Zhang_etal_2006}.

\begin{figure}
  \epsscale{1.15}
  \includegraphics[width=\columnwidth]{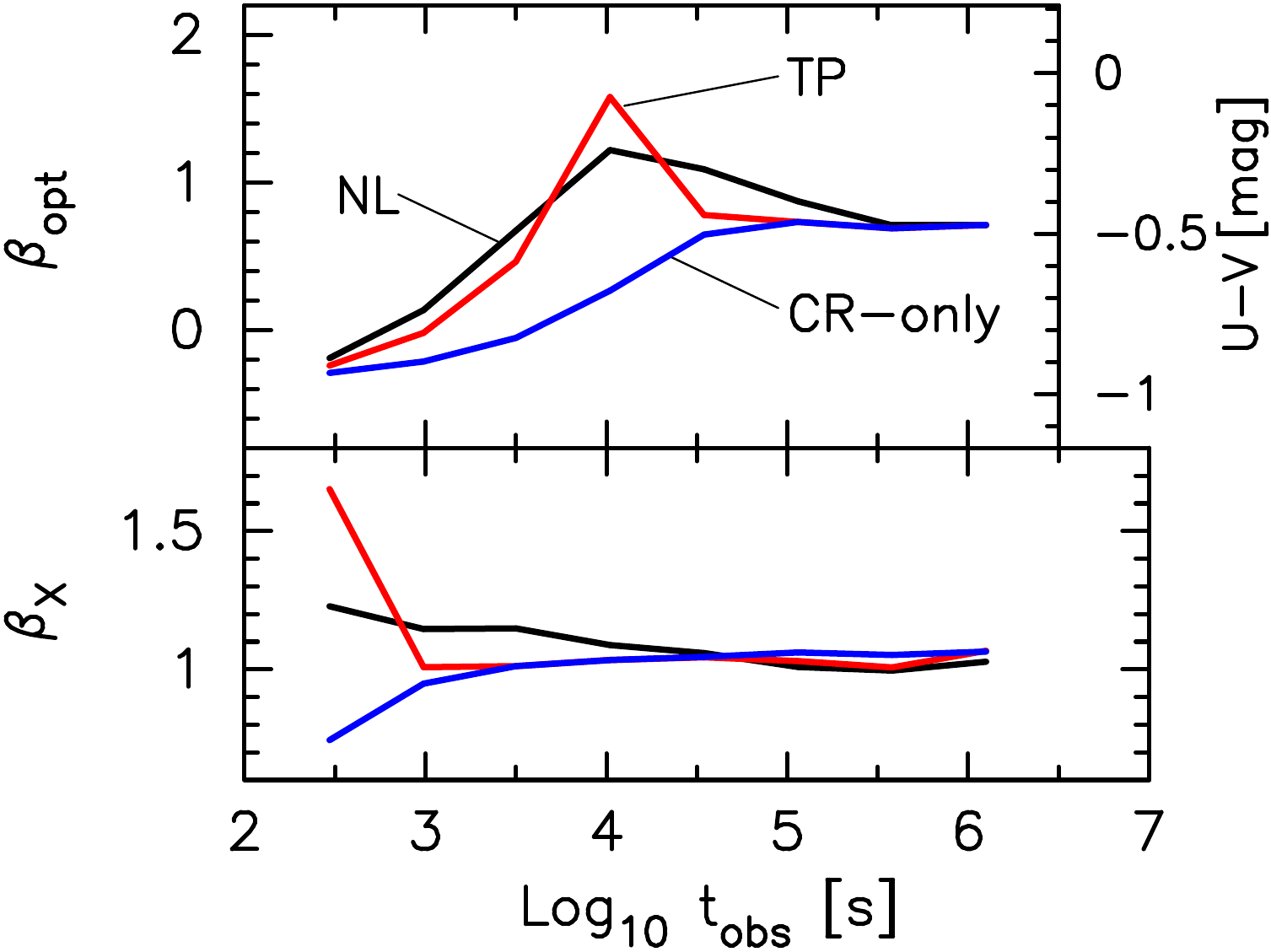}
  \caption{\textit{Top panel}: ratio of flux in $U$ band to flux in $V$ band for the three models, as a function of time; the right-hand $y$ axis shows this value as a difference in magnitudes.  As elsewhere, the CR-only case is plotted with a blue curve, the TP model with red, and the NL model with black.  \textit{Middle panel}: the spectral index $\beta_\mathrm{opt}$ ($F_{\nu} \propto \nu^{-\beta}$) for optical emission from the three models. \textit{Bottom panel}: the X-ray spectral index $\beta_{X}$.}
  \label{fig:color_compare}
\end{figure}

Looking at light curves alone---Figure~\ref{fig:optxray_compare}---turns out to be a poor way to distinguish between models with and without efficient particle acceleration.  Though it does test whether there is a population of shock-heated thermal particles that did not enter the acceleration process, the dependence on efficiency manifests most clearly in optical energies, where there is not a standard template to compare against.  With access to spectra at various times, the differences become much clearer.  In Figure~\ref{fig:color_compare} we show three different measures of the spectral evolution of the afterglow at various times.  The top panel shows the optical spectral index $\beta_\mathrm{opt}$ and the $U-V$ color of all three models.  In the bottom panel we plot the X-ray spectral index, $\beta_{X}$.

We note first that the optical and X-ray spectral evolutions behave in virtually identical manners, barring a time offset.  Our Monte Carlo simulations did not cover high enough shock speeds (i.e., early enough times) to capture the full X-ray spectral evolution, but all predictions our model makes for optical spectra apply to X-ray as well---and conversely, our model predicts that observations in X-ray would be echoed at later times in the optical regime.  We will focus our discussion of spectral evolution on optical, and note that \citet{GianniosSpitkovsky2009} obtained almost the same results in X-ray with their analytical model.

The optical spectrum of the CR-only case shows a monotonic evolution from hard, $\beta_\mathrm{opt} = -0.3$ (before the passage of the minimum energy) to soft, $\beta_\mathrm{opt} = 0.7$ (after the passage).  While the TP and NL models share these end points, the inclusion of thermal particles significantly alters the behavior in the time between.  The dropoff in particle number at higher energies than the thermal peak causes a temporary softening of the optical spectra, before they harden again to their late-time values.  This is a robust prediction of any model that includes a thermal population of particles in addition to the nonthermal shock-accelerated population \citep{GianniosSpitkovsky2009}.

Importantly, the height and width of the peak in $\beta$ appear to be correlated with the efficiency of particle acceleration, since the shock precursors in the NL model smooth out the transition from emission due to the thermal peak and emission due to the nonthermal tail.  The NL model shows a broad peak in $\beta_\mathrm{opt}$ with a maximum value of $\beta_\mathrm{opt} = 1.2$.  The TP model is much more sharply peaked, and exceeds $\beta_\mathrm{opt} = 1.6$ (the true peak may be even higher, as the curve presented in Figure~\ref{fig:color_compare} is necessarily discretized to reflect our limited number of time steps).  While some of this may be due to the difference in particle count \citep[and therefore the size of the dropoff between thermal and nonthermal populations; cf. Figure~8 in][]{GianniosSpitkovsky2009}, our models show that the peak in the spectral evolution curves occur at the same time instead of being dependent on the fraction of particles in the nonthermal tail.  As well, the NL model both softens and hardens more slowly than does the TP model, again due to the bridge between the thermal peak and high-energy tail of electrons.

Numerous bursts show this behavior in the X-ray band \citep{ZLZ2007}, though we are unaware of bursts showing similar evolution at optical wavelengths.  This is very likely due to our parametrization of conditions around the subshock itself.  The behavior shown in Figure~\ref{fig:color_compare} is caused almost entirely by the temporal evolution of the thermal population, which did not enter the acceleration process and so was energized only once, at the first shock crossing.  Accurate treatment of that crossing is vital to predicting the behavior of thermal particles at late times.  Most importantly, our assumption that $f_\mathrm{trans}$ was constant throughout the afterglow does not match theoretical predictions \citep{LemoinePelletier2011}.  A reduction in $f_\mathrm{trans}$ would necessarily affect the energy of the thermal peak.  Per PIC simulations, such a change would be driven by decay of the intensely amplified magnetic field that mediates the energy transfer.  This would depress the energy of the thermal peak, and the associated peak in synchrotron emission, in a time-dependent manner at odds with our simple assumption.

The X-ray spectral indices of the TP and NL models show fluctuations at the latest few time steps.  This is caused by interaction between the thermal peak of the SSC spectrum and the high-energy tail of the synchrotron spectrum, as discussed around Figure~\ref{fig:phot_compare}; the CR-only case, with its suppressed SSC emission, shows almost no variation during this period.

\begin{figure}
  \epsscale{1.15}
  \includegraphics[width=\columnwidth]{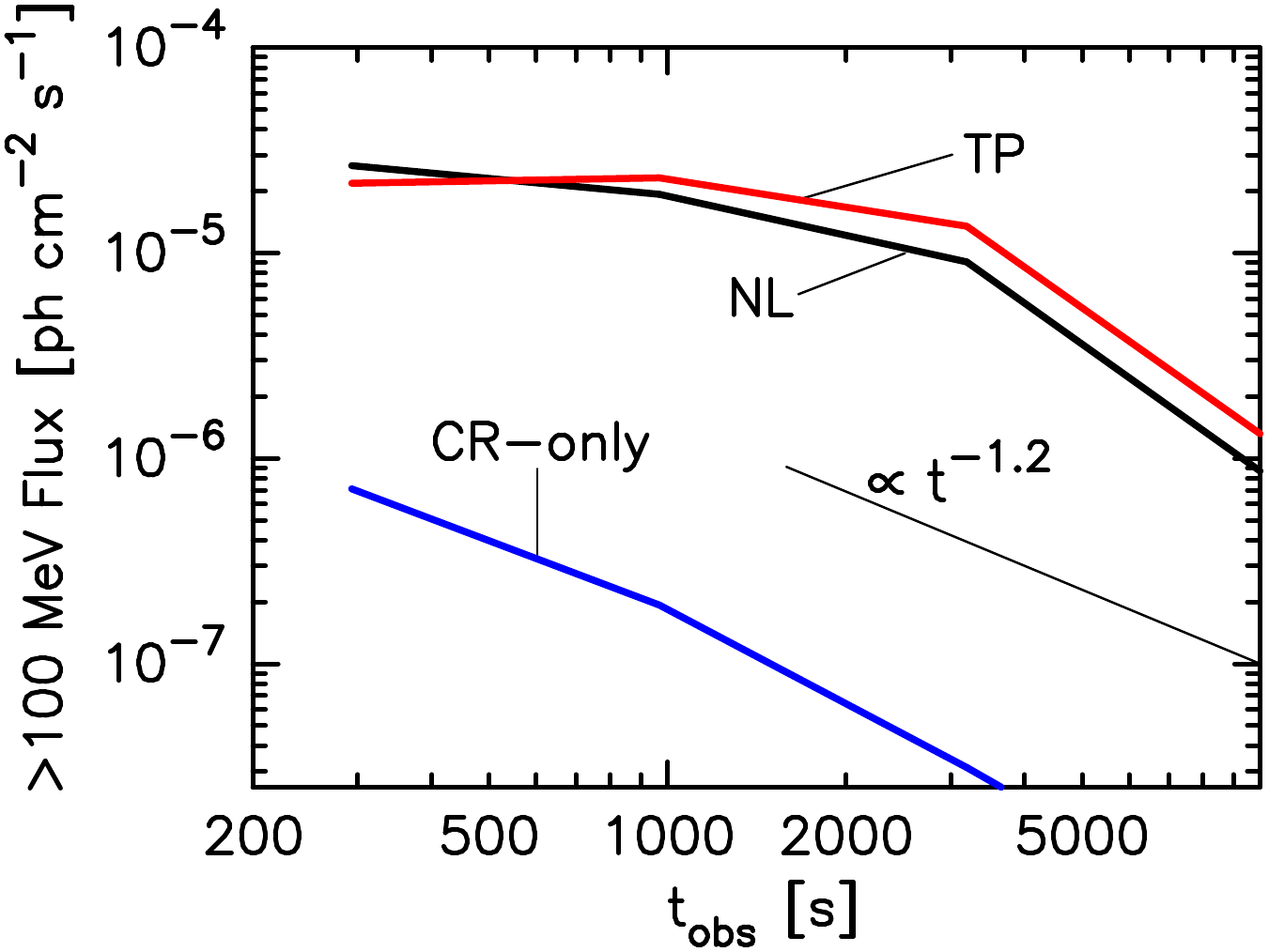}
  \caption{Total photon flux above 100 MeV early in the afterglow.  The high-energy light curve of the TP model is marked with a red line, that of the CR-only case with a blue line, and the light curve of the NL model with a black line.  The thin black line at lower right shows a decay index of -1.2.}
  \label{fig:GeV_compare}
\end{figure}

Both the TP and NL models predict a great deal of emission at $\gamma$-ray energies, even after EBL absorption, as can be clearly seen in Figure~\ref{fig:phot_compare}.  The lack of thermal particles in the CR-only case strongly depresses SSC emission relative to the TP and NL models.  Early in the afterglow there are many observations against which our model may be compared; see \citet{Panaitescu2016} for a discussion of the two dozen bright bursts that have been observed over a significant length of time.  In that paper, a distinction was drawn between bright-soft bursts (whose initial flux above 100~MeV is high, but which decay rapidly) and faint-hard bursts.  The temporal decay chosen to demarcate the two populations was $t^{-1.2}$, corresponding to a trough in the distribution of observed decay indices.

We plot the expected light curves (in terms of photon flux rather than energy flux) from the CR-only case and the TP \& NL models at photon energies higher than 100 MeV in Figure~\ref{fig:GeV_compare}.  We note that the CR-only case follows the $\alpha = 1.2$ guide line very closely.  This is entirely sensible, given its origin as high-energy synchrotron emission, $\nu > \mathrm{max}(\nu_{m},\nu_{c})$, from a power-law distribution of particles accelerated by a relativistic shock.  In the TP and NL shocks, the thermal peak of the SSC spectrum exceeds the synchrotron spectrum at 100~MeV in energy until well into the afterglow (see Figure~\ref{fig:phot_compare}).  The passage of the SSC thermal peak through this energy range causes a plateau in emission, just as occurred in both X-ray and optical in Figure~\ref{fig:optxray_compare}.  As thermal SSC emission gives way to SSC emission from photons originating with the non-thermal population, the GeV light curve steepens.  A well-covered GeV light curve that showed a break should be a strong indicator of the presence of thermal particles producing SSC emission at these energies.

As with the optical in Figure~\ref{fig:optxray_compare}, the presence or absence of thermal particles makes a substantial difference in the flux above 100~MeV.  Though the difference is initially just a factor of 30, the plateau in the TP \& NL models' emission caused by the the thermal peak rapidly increases the disparity.  Afterglow parameters ($\epse$, $\epsB$, $E_\mathrm{iso}$, fraction of particles in the nonthermal tail, etc.) could be adjusted to bring the CR-only case's emission in line with the other two.  We reiterate that thermal particles---which are certain to be present in the shocked plasma---participating in the photon production process yields large gains in emission with no tweaking of parameters needed.

Despite the large factor by which the TP \& NL models are brighter than the CR-only case, both are fainter in the Fermi-LAT band than most of the afterglows with observed GeV emission.  There are reasons to suspect that the model presented here underpredicts the amount of high-energy emission, though.  As discussed elsewhere, we assume a weaker magnetic field near the subshock than is observed in PIC simulations \citep[where $\epsB$ can locally exceed $0.1$, as in][]{Spitkovsky2008,SSA2013}.  The higher fields would allow for more, and higher-energy, synchrotron photons for a given electron.  This increase in synchrotron photons would, of course, increase the observed photon flux; it would also increase the ratio of synchrotron to SSC photons, allowing the NL and TP models to decay with $\alpha = 1.2$ for a longer period before SSC emission is significant at 100~MeV energies.  On the other hand, an increase in synchrotron and SSC emission would both be associated with stronger cooling, as discussed in Section~\ref{sub:observables}, which would reduce the emission at the highest energies.  The degree to which these competing effects balance each other out will be explored in future extensions of this model.

The strongest limits we are aware of on later-time gamma-ray flux from GRB afterglows are presented in \citet{Aliu_etal_2014}, who observed GRB 130427A on three consecutive nights after the GRB ($\approx$~20~hr, $\approx$~44~hr, and $\approx$~68~hr after the trigger).  Their non-detections placed upper limits (respectively) of 9, 7, and $3\xx{-12}$~erg~cm$^{-2}$~s$^{-1}$ on flux in the VERITAS range, $E_{\gamma} > 100$~GeV.  The first observation, at 20 hours after the GRB, would occur between time steps 5 and 6 (green and yellow curves in Figure~\ref{fig:phot_compare}) of our model afterglows.  By this point, even the TP and NL models are below the limits established by \citet{Aliu_etal_2014}, so the VERITAS non-detections are consistent with our results.

\begin{figure}
  \epsscale{1.15}
  \includegraphics[width=\columnwidth]{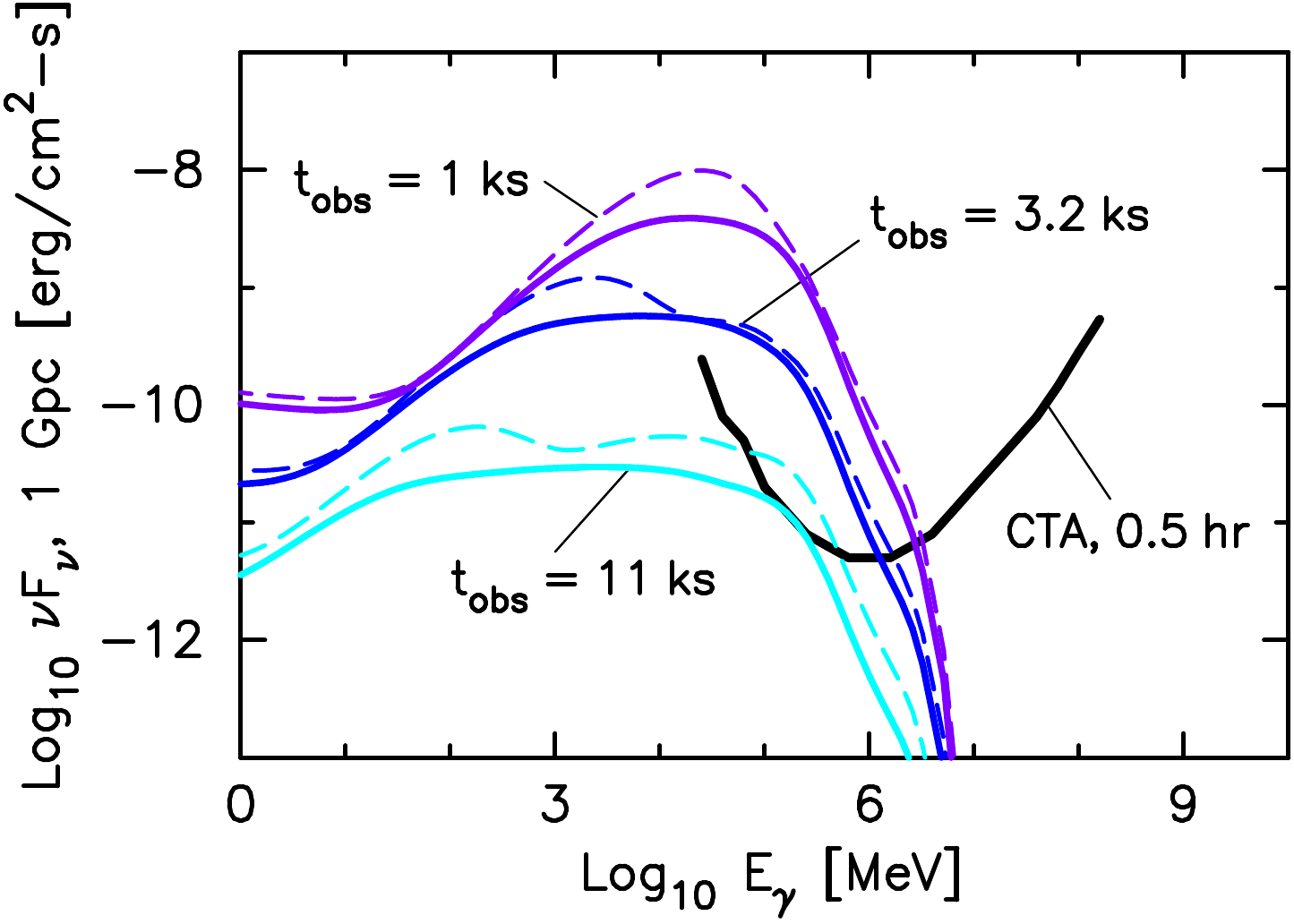}
  \caption{Comparison of high-energy photon emission from the TP and NL models at early times.  Dashed lines come from the TP model, and solid lines from the NL model.  Colors refer to time steps as in Table~\ref{tab:afterglow_common}.  The black curve at the right is the CTA North sensitivity curve for a half-hour observation.}
  \label{fig:phot_CTA_compare}
\end{figure}

In spite of the fact that the TP and NL models predict non-observations by VERITAS and minimally-significant observations by Fermi, they are well above the sensitivity curves for the Cherenkov Telescope Array (CTA).  In Figure~\ref{fig:phot_CTA_compare} we repeat the second, third, and fourth time steps from the bottom two panels of Figure~\ref{fig:phot_compare}, overlaid to provide a more direct comparison of the changes due to acceleration efficiency.  In addition, we plot the expected sensitivity curve for a half-hour observation by the CTA North site.\footnote{The CTA curve uses data taken from https://portal.cta-observatory.org/Pages/CTA-Performance.aspx .}  It is clear from the figure that our models with thermal particles are well above the detection threshold for the array, and they remain above the sensitivity curve for approximately 10~ks after the initial trigger.  If, as discussed above, our model underpredicts the GeV flux, GRB afterglows would be detectable even longer and/or to greater distances, increasing the frequency at which CTA should observe these events.

\section{Conclusions}
\label{sec:conclusions}

In this work we have presented an evolutionary model for GRB afterglows that takes into account the nonlinear (NL) interaction between shocks and the cosmic rays they accelerate.  Key assumptions of the Monte Carlo model were that acceleration may be efficient (moving beyond the test-particle, TP, regime) and that the particles scattering in the shock structure undergo pitch-angle scattering regardless of position within the shock structure.  We additionally assumed an upstream magnetic field strong enough that it would reach a downstream energy density fraction of $\epsB \sim 10^{-3}$ through compression only.  In physical shocks these extremely high magnetic fields result from amplification of the upstream field through turbulent instabilities.

The key result of this paper is when thermal\footnote{See footnote~\ref{ftnt:thermal_def}.} particles---which were heated by their initial shock crossing but not injected into the acceleration process---participate in photon emission, the resultant spectra and light curves depart substantially from the standard synchrotron model.  These changes stem from the shape of the electron spectrum (Figure~\ref{fig:spectrum_compare}): efficient NL shocks form a smooth (but steep) bridge connecting the thermal peak to the high-energy non-thermal tail, while the transition is much sharper in less efficient TP shocks.  Both models, however, show a drop in number from the thermal peak to the non-thermal tail; this stands in sharp contrast to the CR-only case for afterglows, which assumes that the spectrum of radiating electrons is a pure power law.

Regardless of how efficiently shocks inject particles for acceleration, emission due to uninjected thermal particles enhances the predicted flux at optical (Figure~\ref{fig:optxray_compare}) and GeV (Figure~\ref{fig:GeV_compare}) energies; at X-ray energies (again, Figure~\ref{fig:optxray_compare}) emission is almost always due to synchrotron production of high-energy electrons, and so the three models (NL, TP, and CR-only) are more similar.  The increase in flux at GeV energies greatly improves the chances for detection by future $\gamma$-ray observatories like CTA (Figure~\ref{fig:phot_CTA_compare}).  The GeV spectra of the TP and NL models can also explain both the energies of the highest-energy photons associated with GRBs (such as the 94~GeV photon associated with GRB 130427A) and the delay between the trigger and photon detection.

In addition to the differences in light curves, including thermal particles or efficient acceleration causes spectral evolution.  When a thermal population is included, the optical and X-ray spectral indices change from hard, to soft, to hard again (Figure~\ref{fig:color_compare}).  The height and width of this peak, or indeed whether there is a peak at all, can be used to gauge the efficiency of energy transfer from ions to electrons (i.e., the existence of a relativistic thermal population downstream) and the efficiency of particle injection into the acceleration process (whether the shocks are closer to the test-particle or nonlinear regimes).

The Monte Carlo approach used in this work is versatile, but must parametrize many important aspects of the shocks.  We rely on PIC simulations and analytical treatments to inform our model of the subshock, where the turbulent field is strongest and velocity gradients are steepest.  Both PIC simulations and analytical work are sorely needed in the trans-relativistic regime ($\gamma_{0} \lsim 3$), where energy transfer from ions to electrons is not yet established.

\acknowledgments 
Part of this work was performed at the Aspen Center for Physics, which is supported by National Science Foundation grant PHY-1066293.  We would like to thank Susumu Inoue and Alexei Pozanenko for their numerous helpful comments and discussions, which greatly improved the quality and clarity of the paper.  SN acknowledges support from the Japan Society for the Promotion of Science (Grant No. 26287056); the Mitsubishi Foundation Research Grants in the Natural Sciences (2016-2017); the Competitive Program for Creative Science and Technology Pioneering Projects (iTHES, E3), RIKEN; and the Start-up Budget for an Associate Chief Scientist, RIKEN.

\bibliographystyle{aa} 
\bibliography{dcw}

\end{document}